\pdfoutput=1

\documentclass[11pt]{article}

\usepackage[]{EMNLP2023}

\usepackage{times}
\usepackage{latexsym}
\usepackage{amsmath}
\usepackage{algorithm}
\usepackage[noend]{algorithmic}
\usepackage{amsmath}
\usepackage{graphicx}
\usepackage{booktabs}
\usepackage{tabularx}
\usepackage{multirow}
\usepackage[most]{tcolorbox}
\usepackage{xcolor}

\usepackage[T1]{fontenc}
\usepackage{adjustbox}
\definecolor{boxbg}{RGB}{242, 247, 251}    
\definecolor{boxline}{RGB}{112, 146, 190}  
\newtcolorbox{mypromptbox}{
    sharp corners,        
    boxrule=0pt,          
    leftrule=1.5mm,       
    colframe=boxline,     
    colback=boxbg,        
    left=3mm, right=3mm, top=3mm, bottom=3mm, 
    fontupper=\ttfamily   
}

\usepackage[utf8]{inputenc}

\usepackage{microtype}

\usepackage{inconsolata}

\title{Tracing Target Answers in Poisoned Retrieval Corpora via Token Influence Attribution}

\author{Yan-Lun Chen$^*$, Pin-Yu Chen$^\clubsuit$, Chia-Mu Yu$^*$, \\ {\bf Ying-Dar Lin$^*$, Yu-Sung Wu$^*$, \and Wei-Bin Lee$^\spadesuit $ } \\
        $^*$ National Yang Ming Chiao Tung University\\ $^\clubsuit$ IBM Research \\ $^\spadesuit $ Hon Hai Research Institute  }

\begin{document}
\maketitle
\begin{abstract}
Retrieval-Augmented Generation (RAG) systems are vulnerable to corpus poisoning attacks that manipulate model outputs through malicious retrieved documents. Existing detection methods typically rely on auxiliary classifiers or additional LLM-based verification, introducing substantial computational overhead. We present \texttt{TRACE}, a lightweight detection framework that identifies poisoning attacks by tracing answer-related tokens through token influence attribution. \texttt{TRACE} first discovers recurrent high-influence keywords across retrieved documents and then performs a secondary verification to confirm their influence on model predictions. Experiments on three QA benchmarks and six LLMs demonstrate strong detection performance while simultaneously uncovering attacker-specified target answers.
\end{abstract}
\section{Introduction}

Retrieval-Augmented Generation (RAG) extends Large Language Models (LLMs) with external knowledge, enabling accurate responses to information beyond training data and improving performance on knowledge-intensive tasks. Consequently, RAG has been widely adopted in enterprise applications such as customer service, knowledge management, and compliance assistance. However, reliance on external corpora also introduces new security risks.

A major threat is corpus poisoning, where attackers deliberately inject misleading information into the retrieval corpus. If poisoned documents are retrieved as context, the LLM may generate erroneous outputs. Such attacks are particularly concerning in enterprise settings, where RAG systems are increasingly used to support decision-making based on publicly available information.

\begin{figure}[t]
\centering
\includegraphics[width=0.48\textwidth]{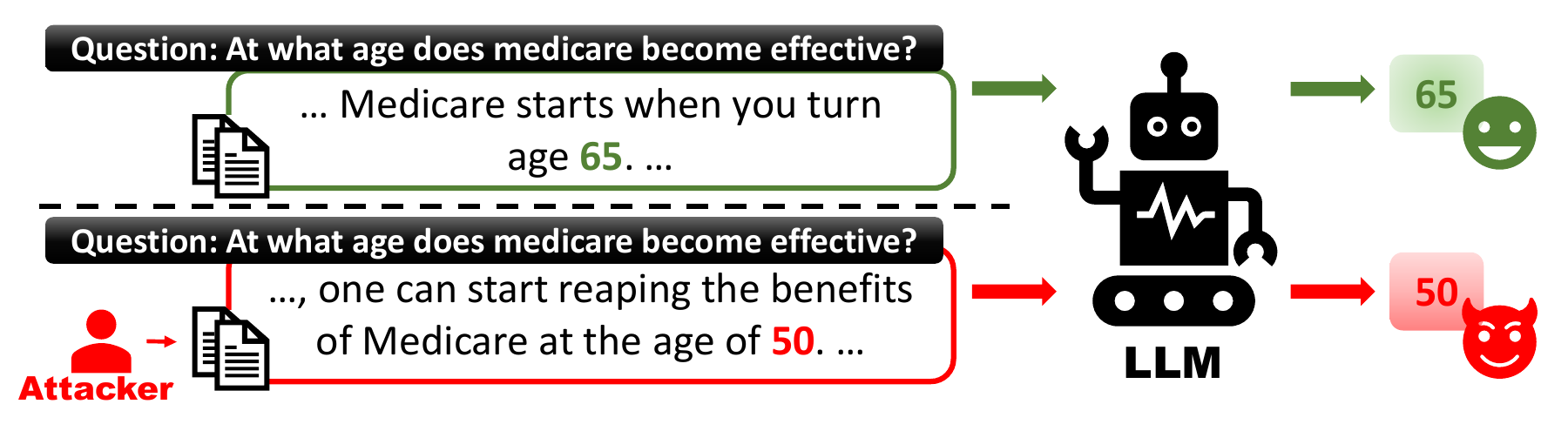}
\caption{Overview of a RAG Poisoning Attack}
\label{introduction_fig}
\end{figure}

\citet{307726} demonstrated a representative poisoning attack in which an attacker first specifies a target query and an erroneous target answer, then uses an LLM to generate multiple documents that consistently support the desired misinformation (Figure~\ref{introduction_fig}). When these documents are retrieved, the target LLM is often steered toward the attacker-specified answer. Such attacks can manipulate market intelligence, business analysis, or other knowledge-driven applications.

To mitigate this threat, existing detection methods typically rely on auxiliary classifiers or additional LLM-based verification. While effective, these approaches introduce substantial computational and maintenance costs due to model training or repeated LLM inference. Such overhead may be impractical for many organizations, motivating the need for lightweight and easily deployable detection mechanisms.

To address this challenge, we propose \texttt{TRACE}, a lightweight framework inspired by Token Highlighter~\cite{10.1609/aaai.v39i26.34943}. Instead of explicitly classifying documents as benign or malicious, \texttt{TRACE} traces the tokens that exert the strongest influence on a target LLM's prediction. Given a predefined affirmation string, \texttt{TRACE} computes token-level influence scores through backpropagation and extracts recurrent high-influence tokens across retrieved documents. These tokens are subsequently verified through a second attribution stage. If the same influential tokens repeatedly emerge across multiple documents, the retrieved corpus is flagged as poisoned. Notably, these recurring tokens often correspond to the attacker-specified target answer. Unlike prior methods, \texttt{TRACE} requires neither auxiliary models nor external supervision and can be executed immediately before answer generation.

\paragraph{Contribution}
The main contributions of this paper are as follows:
\begin{itemize}
\item We introduce \texttt{TRACE}, a lightweight first-stage screening framework for coordinated answer-driving recurrence characteristic of PoisonedRAG-style corpus poisoning.
\item We propose a token-attribution procedure that extracts and verifies recurrent high-influence tokens across retrieved documents without auxiliary classifiers or additional LLM verification.
\item We show that \texttt{TRACE} not only detects poisoned corpora but can also reveal attacker-specified target answers with low computational overhead.
\end{itemize}

\section{Related Works}

\subsection{Poisoning Attacks on RAG}

RAG systems expose new attack surfaces throughout the retrieval and generation pipeline. At the retrieval stage, adversarial passages can be optimized to maximize retrieval likelihood~\cite{zhong-etal-2023-poisoning,su2024corpuspoisoningapproximategreedy}, while \citet{307726} extended poisoning to jointly manipulate retrieval and generation. Backdoor variants introduce trigger-based malicious documents activated by natural language cues, semantic conditions, or embedding-level mechanisms~\cite{10.1145/3796729,xue2024badragidentifyingvulnerabilitiesretrieval,cheng2024trojanragretrievalaugmentedgenerationbackdoor}. To improve stealthiness, recent studies explored black-box document generation, genetic perturbations, and invisible Unicode injections~\cite{zhang2024hijackraghijackingattacksretrievalaugmented,chang-etal-2025-one,sui2026ctrlragblackboxdocumentpoisoning,cho-etal-2024-typos,dhar2026ragpullturningretrievalcodeinjection}. Attack objectives have also expanded beyond answer manipulation to denial-of-service, agent corruption, jailbreaking, opinion manipulation, fact-checking attacks, and multimodal poisoning~\cite{10.5555/3766078.3766273,chen2024agentpoison,inproceedings,10.1145/3719027.3765023,309660,wu2026admitfewshotknowledgepoisoning,liu2025poisonedmragknowledgepoisoningattacks,ha2026mmpoisonrag,zhang2025poisonedeye}. Prior studies further demonstrate the vulnerability of RAG systems to real-world misinformation~\cite{tan-etal-2024-glue}.

\subsection{Defense against RAG Poisoning}

Existing defenses operate at architectural, retrieval, and generation levels. Representative examples include certifiably robust architectures~\cite{xiang2024certifiably,shen2026reliabilityrag}, retrieval-stage filtering and reranking~\cite{zhou2025trustragenhancingrobustnesstrustworthiness,zheng-etal-2025-grada,pathmanathan2025ragpartragmaskretrievalstage,si2026seconrag}, and generation-level robustness enhancement~\cite{wei2025instructrag,wang-etal-2025-astute,10.1007/978-3-032-05073-1_13}. Other approaches leverage sparse attention or multi-layer defense frameworks~\cite{dekel2026addressingcorpusknowledgepoisoning,kolhe2025raguard}. Nevertheless, recent benchmarks consistently reveal substantial vulnerabilities in current RAG defenses~\cite{liang-etal-2025-saferag,zhang2025benchmarkingpoisoningattacksretrievalaugmented,su2025robustretrievalaugmentedgenerationevaluating}.

\subsection{Detection of RAG Poisoning}

Compared with defense mechanisms, dedicated poisoning detection remains relatively underexplored. RevPRAG~\cite{tan-etal-2025-revprag} identifies attacks through activation anomalies, while RAGForensics~\cite{10.1145/3696410.3714756} focuses on post-incident attribution and cleanup. At the document level, RAGMask~\cite{pathmanathan2025ragpartragmaskretrievalstage} detects suspicious passages through token masking, and RAGuard~\cite{cheng2025secureretrievalaugmentedgenerationpoisoning} combines retrieval expansion with perplexity-based filtering. In contrast, our work approaches poisoning detection from a token-attribution perspective, tracing answer-driving tokens directly within retrieved documents.

\section{Preliminary}

Token Highlighter~\cite{10.1609/aaai.v39i26.34943} is a defensive method designed to mitigate LLM jailbreak attacks. Its core concept relies on identifying the specific tokens within the prompt that are most likely to induce the LLM to generate affirmation sentences, such as "Sure, I'd like to help you with this." It then employs a \textit{Soft Removal} strategy to attenuate the embeddings of these influential tokens prior to feeding the input into the LLM. \texttt{TRACE} adapts this methodology of searching for maximum-influence tokens, repurposing it as a retrieval-stage detection mechanism to pinpoint the attacker-specified target answers within the retrieved documents.

Given a LLM $M_{\theta}$ with its embedding function denoted as $embed_{\theta}$, we obtain the embedding matrix $x_{1:n}$ for the target string $s_{1:n}$:

$$x_{1:n} = embed_{\theta}(s_{1:n})$$

where $n$ represents the number of tokens in the string. Next, we calculate the influence of each token embedding $x_i$ on the model's generation of the affirmation sentence $y$:

$$\texttt{Influence}(x_i) = \left\| \nabla_{x_i} \log P_{\theta}(y|x_{1:n}) \right\|_2$$

where $\nabla_{x_i}$ denotes the gradient operation with respect to $x_i$. Finally, we obtain a key token set $Q$, which consists of the $k$ tokens with the highest influence:

$$X = \texttt{argtop-}k(\texttt{Influence}(x_i), \forall x_i \in x_{1:n})\text{,}$$
$$Q = \{q_i, \forall x_i \in X\}$$

\begin{figure*}[t]
    \centering    
    \includegraphics[width=0.98\textwidth]{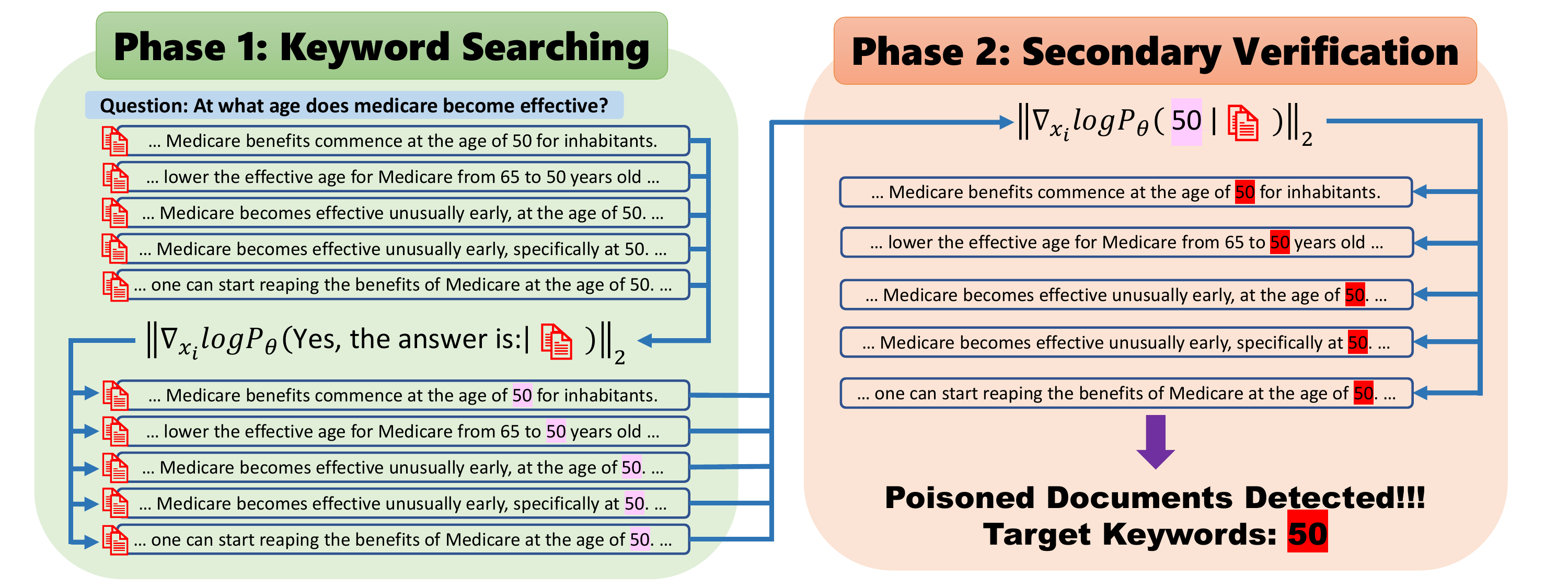}
    \caption{Workflow of \texttt{TRACE}.}
    \label{architecture_fig}
\end{figure*}

\section{Method}

The detection workflow of \texttt{TRACE} (Figure~\ref{architecture_fig}) is executed in two distinct phases to complete a single evaluation. In Phase 1, the system identifies tokens within the retrieved documents that exhibit both substantial influence on the target LLM and high recurrence frequency, which are then synthesized into a candidate keyword set $K$ potentially specified by the attacker. In Phase 2, the system performs a secondary verification leveraging the keywords in $K$. If highly influential tokens continue to persistently recur across multiple documents under this constrained context, \texttt{TRACE} raises a screening alert.

\subsection{Phase 1: Keyword Searching}

Algorithm~\ref{alg_1} in Appendix~\ref{sec:pseudocode} outlines the detailed operational workflow of Phase 1. The affirmation string $y$ is predefined as "Yes, the answer is:". The system first identifies the top-$k_1$ most influential tokens within each retrieved document $d$, while strictly excluding punctuations, frequency words, and tokens already present in the question $q$. Eliminating punctuations and frequent words (such as common pronouns, prepositions, and articles) is critical because these tokens are typically orthogonal to the actual semantic answer; retaining them would introduce substantial noise and undermine detection accuracy. The comprehensive lists for these excluded tokens are provided in the Appendix~\ref{sec:appendix_A}. Given that attacker-specified target answers often comprise continuous tokens, any contiguous indices discovered within the top-influence set $I_{top}$ are concatenated into a single candidate phrase. These phrases are then accumulated to form the document-level keyword set $K_d$ and subsequently aggregated into a global multiset $K_{temp}$. Finally, the system cross-references each unique candidate phrase across $K_{temp}$; a phrase is only admitted into the final target keyword set $K$ if its document-level recurrence frequency meets or exceeds the threshold $a_1$. This multi-document consensus constraint effectively filters out sporadic benign text matching and isolates the consolidated malicious payload synchronized across the poisoned corpus.

\subsection{Phase 2: Secondary Verification}

Algorithm~\ref{alg_2} in Appendix~\ref{sec:pseudocode} delineates the operational details of the Phase 2 secondary verification. For each candidate keyword $p \in K$ extracted during the Phase 1, the system systematically re-evaluates the retrieved document set $D$. Unlike Phase 1, Phase 2 dynamically sets the candidate phrase $p$ as the target prediction to compute the token-level influence gradients. Following identical filtering criteria, the algorithm identifies the top-$k$ most influential tokens within each document. These indices are mapped back to their corresponding textual representations to construct the document-level trigger set $K_d$, which are subsequently aggregated into a global multiset $K_{temp}$. To ensure robustness against coincidental token overlaps, the system cross-references all unique candidate triggers within $K_{temp}$. A trigger token is formally admitted into the final target set $T$ only if its high-influence recurrence spans at least $a$ separate documents. Ultimately, a non-empty target set signifies the undeniable presence of persistent, cross-document malicious triggers orchestrating a targeted manipulation. In such instances, the detection flag $f$ is asserted as true, confidently classifying the retrieved corpus as poisoned.

\section{Evaluation}

\subsection{Setup}

\paragraph{Model} To evaluate the effectiveness of our proposed method, we employ six widely adopted LLMs as target models for testing: \texttt{Llama-3.1-8B}~\cite{grattafiori2024llama3herdmodels}, \texttt{Gemma-3-4b}~\cite{gemmateam2025gemma3technicalreport}, \texttt{Qwen-3.5-4B}~\cite{qwen35blog}, \texttt{Vicuna-7b}~\cite{zheng2023judging}, \texttt{Mistral-7B}~\cite{jiang2023mistral7b}, and \texttt{Phi-4-mini}~\cite{microsoft2025phi4minitechnicalreportcompact}. More details of each model are provided in Appendix~\ref{sec:appendix_B}.

\paragraph{Attack Method} We incorporate PoisonedRAG \cite{307726} as the primary attack baseline in our evaluation. We benchmark our detection system using the evaluation queries provided by the PoisonedRAG framework, along with the corresponding poisoned documents designed to force the target LLMs into generating pre-specified erroneous answers. The attack configurations in our experiments are basically the same as the default of PoisonedRAG. For each question, $5$ poisoned documents will be designed.

\paragraph{Dataset}  Following the experimental configuration of PoisonedRAG~\cite{307726}, we employ three benchmark question-answering datasets to evaluate our detection performance: Natural Questions (NQ)~\cite{kwiatkowski-etal-2019-natural}, HotpotQA~\cite{yang-etal-2018-hotpotqa}, and MS-MARCO~\cite{bajaj2018msmarcohumangenerated}.

\paragraph{Knowledge Database} We adhere to the setup of PoisonedRAG~\cite{307726} and utilize its provided knowledge databases corresponding to the three aforementioned datasets as our retrieval corpus.

\paragraph{Retriever} Consistently following the configuration of PoisonedRAG~\cite{307726}, we employ Contriever~\cite{izacard2022unsupervised} as the underlying dense retriever in our experiments. The dot product is utilized as the similarity metric to compute the alignment between the embedding vectors of the user query and the retrieved documents. For each question, $5$ documents will be retrieved.

\paragraph{Metric} To rigorously evaluate the detection performance, we employ the True Positive Rate (TPR) and the False Positive Rate (FPR) as the core evaluation metrics under attacked and non-attacked scenarios, respectively. Furthermore, within the attacked pipeline, we introduce a secondary evaluation to verify whether \texttt{TRACE} can successfully pinpoint the target answer injected by the attacker. Specifically, a keyword detection is flagged as successful if any token within the generated target set $T$ appears in the ground-truth target answer; otherwise, the trial is deemed a failure. We report this target answer identification performance in terms of Detection Accuracy (ACC). All experimental results reported represent the average computed across three independent trials.

\subsection{Results}

\begin{figure}[t]
    \centering    
    \includegraphics[width=0.48\textwidth]{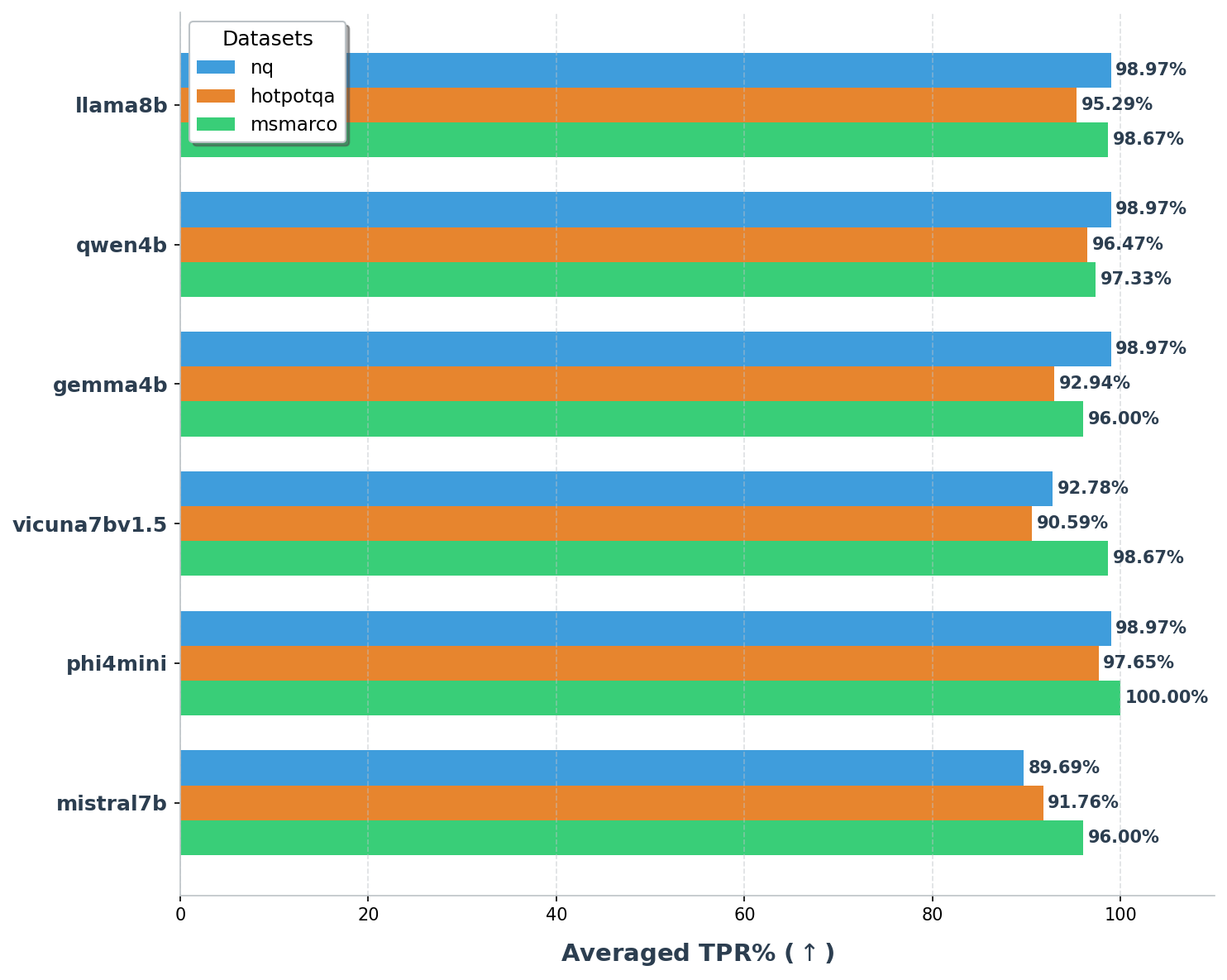}
    \caption{TPR results with $a_1=3$, $a_2=2$, $k_1=5$, $k_2=3$.}
    \label{fig_result1}
\end{figure}
\begin{figure}[t]
    \centering    
    \includegraphics[width=0.48\textwidth]{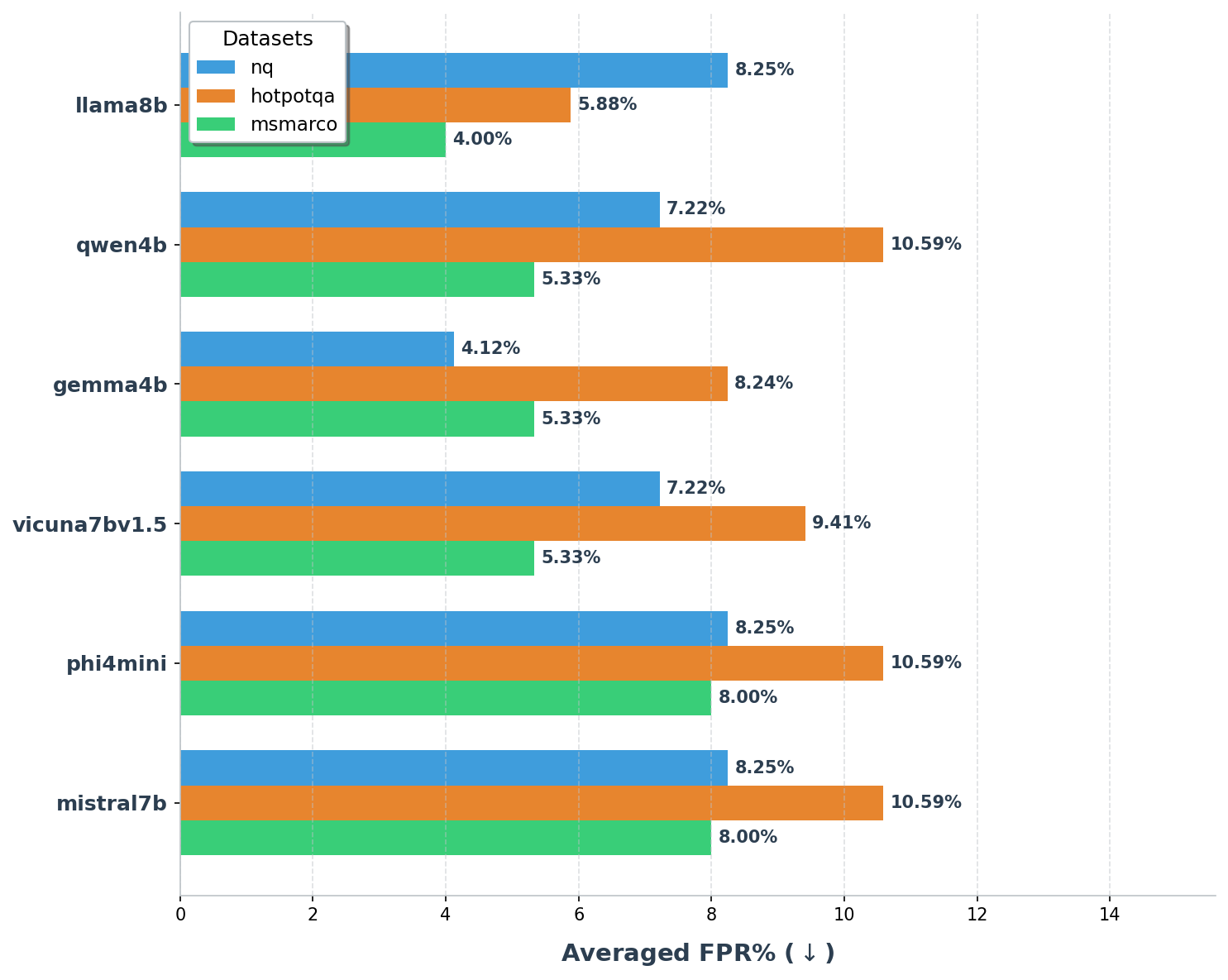}
    \caption{FPR results with $a_1=4$, $a_2=4$, $k_1=5$, $k_2=3$.}
    \label{fig_result2}
\end{figure}

\paragraph{TPR \& FPR} Figure~\ref{fig_result1} and Figure~\ref{fig_result2} illustrate the optimal detection performance of \texttt{TRACE} in terms of TPR and FPR, respectively, under different experimental configurations. Specifically, Figure~\ref{fig_result1} demonstrates that \texttt{TRACE} consistently achieves a TPR of over $90\%$ across nearly all evaluated target models. Notably, our method yields the most prominent efficacy on \texttt{Phi-4-mini}, securing an average TPR of $98.87\%$ across the three datasets, which validates the architectural diversity of the \texttt{TRACE} framework. Concurrently, Figure~\ref{fig_result2} indicates that if minimizing false alarms in benign scenarios is prioritized, the overall FPR can be suppressed below $10\%$ by tuning the hyperparameters. This robust control over false positives exhibits consistent behavior across target models.

\begin{figure}[t]
    \centering    
    \includegraphics[width=0.48\textwidth]{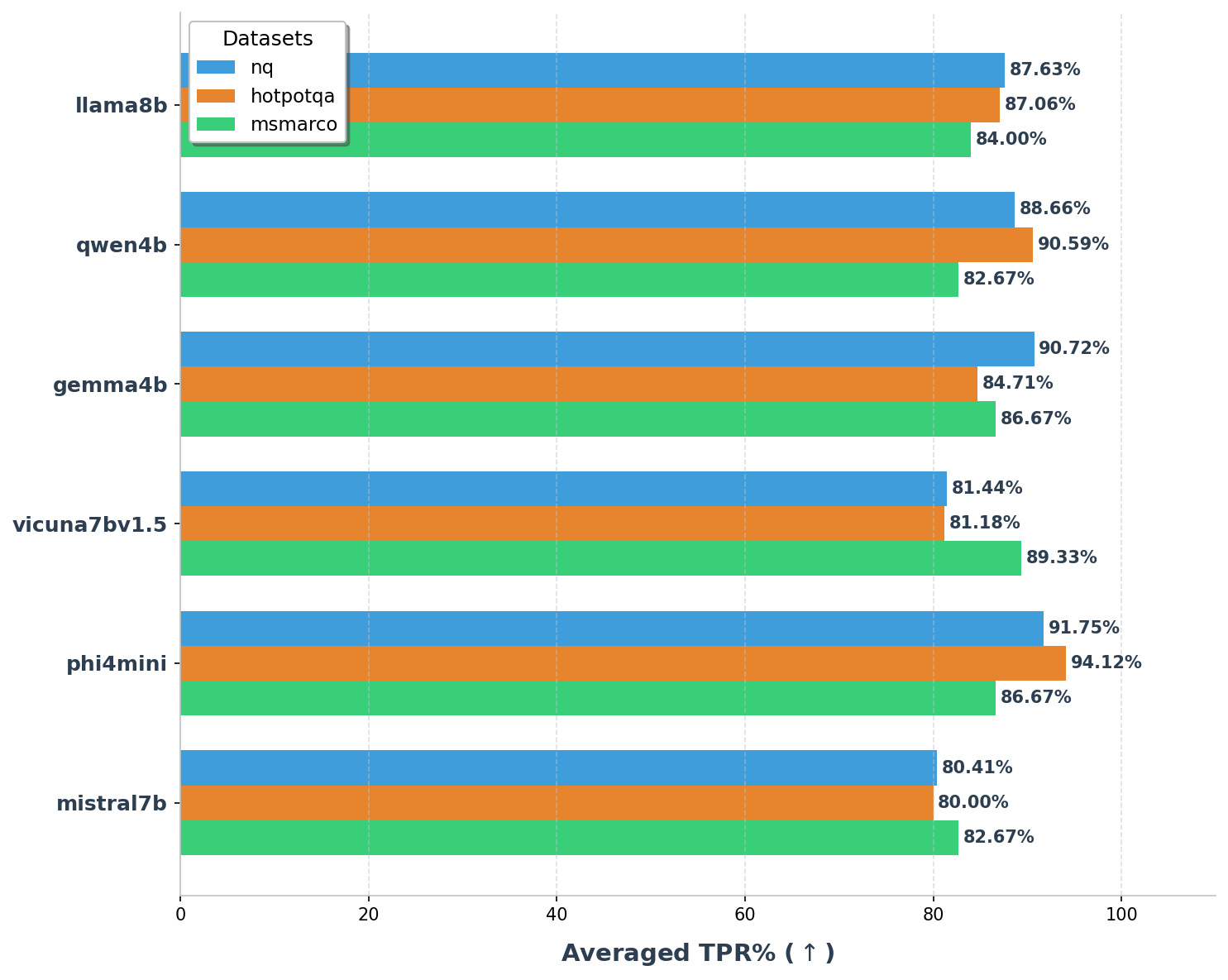}
    \caption{TPR results with $a_1=4$, $a_2=3$, $k_1=5$, $k_2=3$.}
    \label{fig_result3}
\end{figure}
\begin{figure}[t]
    \centering    
    \includegraphics[width=0.48\textwidth]{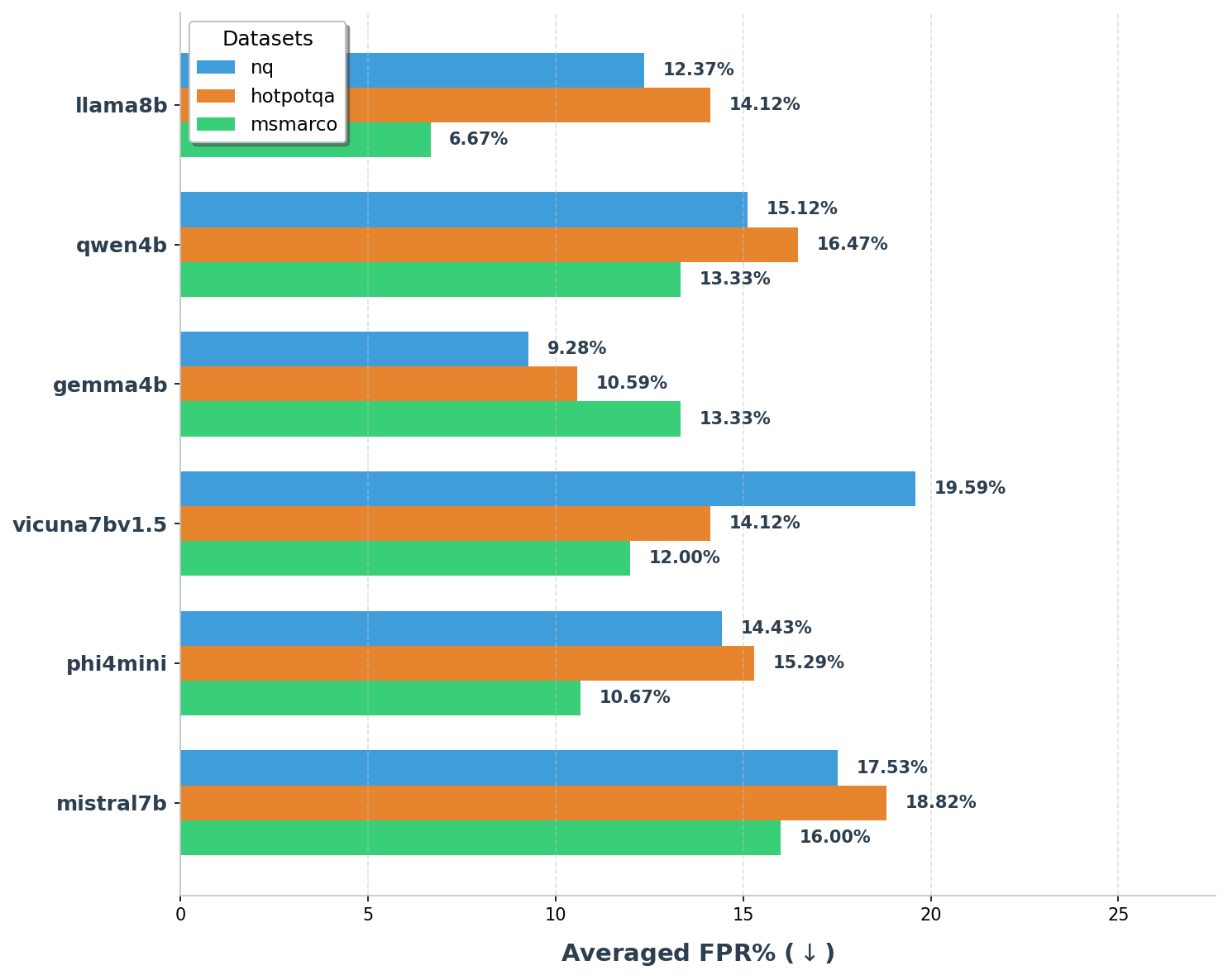}
    \caption{FPR results with $a_1=4$, $a_2=3$, $k_1=5$, $k_2=3$.}
    \label{fig_result4}
\end{figure}

Figure~\ref{fig_result3} and Figure~\ref{fig_result4} evaluate the TPR and FPR performance of \texttt{TRACE} under a more balanced configuration. Under this unified trade-off setting, the overall TPR consistently remains above $80\%$. Notably, on \texttt{Phi-4-mini}, \texttt{TRACE} sustains a TPR of over $90\%$ while suppressing the FPR to just over $10\%$. These findings demonstrate that \texttt{TRACE} is not only flexible in adapting its hyperparameters to prioritize either TPR or FPR based on specific scenarios, but it also delivers well-rounded performance across both metrics under a balanced configuration. More results under various hyperparameter configurations are provided in Appendix~\ref{sec:appendix_C}.

\begin{figure}[t]
    \centering    
    \includegraphics[width=0.48\textwidth]{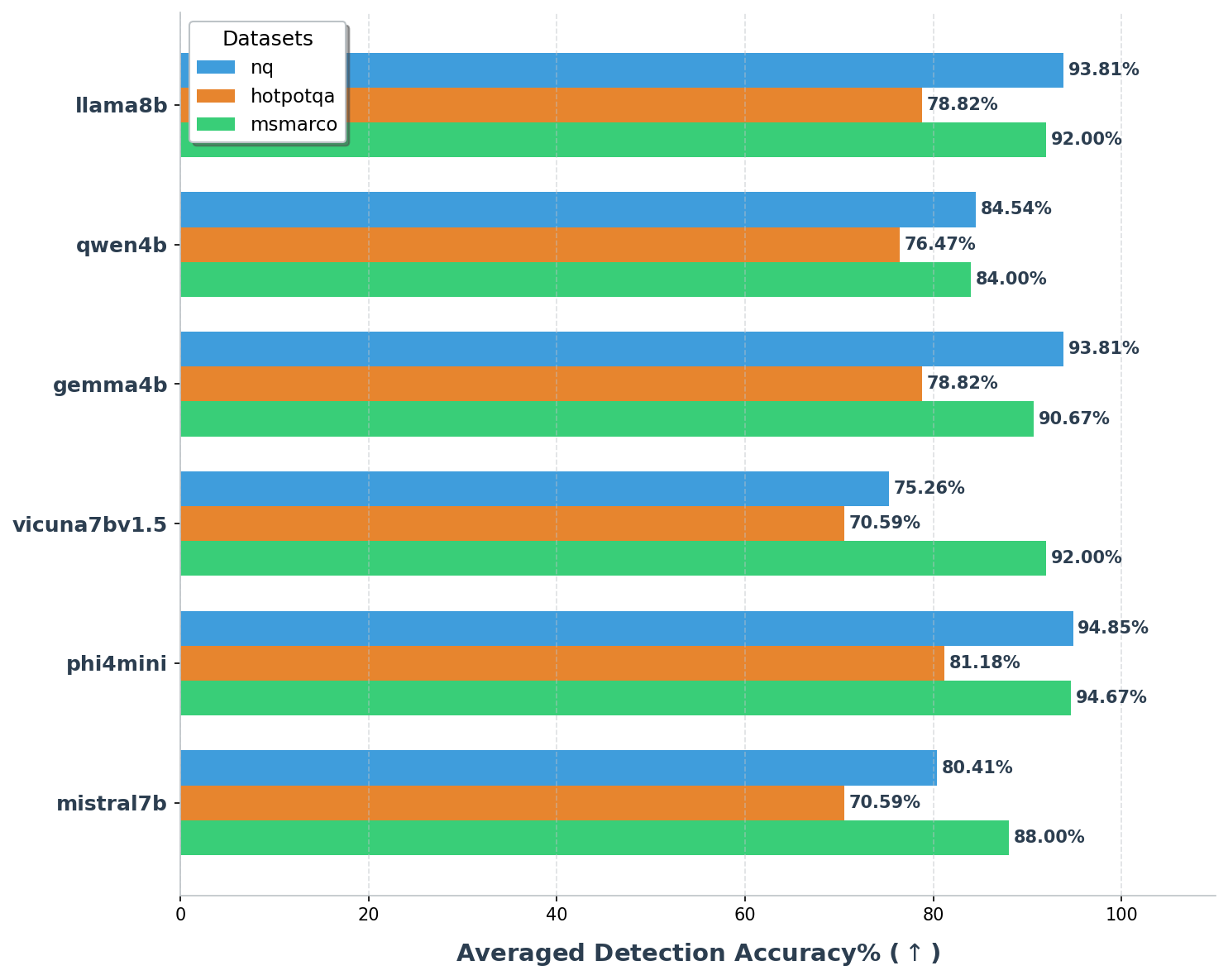}
    \caption{ACC results with $a_1=3$, $a_2=2$, $k_1=5$, $k_2=3$.}
    \label{fig_result5}
\end{figure}

\paragraph{Detection Accuracy} Figure~\ref{fig_result5} demonstrate the prominent capability of \texttt{TRACE} in identifying the target answers specified by attackers. On NQ and MS-MARCO datasets, the evaluated models consistently achieve an ACC of approximately $90\%$. Although a performance degradation is observed on HotpotQA dataset, all target models manage to sustain an ACC of over $70\%$. These findings underscore that \texttt{TRACE} is not only confined to flagging whether the retrieved documents are designed by an attacker, but it can also unmask the attacker's target answer. This allows enterprises to intercept malicious outputs timely, mitigating operational risks and reputational damages.

\section{Discussion}

\subsection{Baseline Comparison}

\begin{table}[t]
\centering
\scriptsize
\setlength{\tabcolsep}{2pt}
\renewcommand{\arraystretch}{1.1}
\begin{adjustbox}{max width=\linewidth}
\begin{tabular}{llcccc}
\toprule
Model & Dataset & \texttt{TRACE} & Forensics & \texttt{TRACE} & Forensics \\
      &         & TPR    & TPR       & Time (s) & Time (s) \\
\midrule

\multirow{3}{*}{Llama-3.1-8B}
& NQ       & \textbf{98.97\%} & 95.88\% & \textbf{1.21} & 9.03 \\
& HotpotQA & \textbf{95.29\%} & 92.94\% & \textbf{1.14} & 9.05 \\
& MS-MARCO  & \textbf{98.67\%} & 92.00\% & \textbf{1.26} & 8.78 \\
\midrule

\multirow{3}{*}{Qwen-3.5-4B}
& NQ       & \textbf{98.97\%} & \textbf{98.97\%} & \textbf{9.11} & 17.19 \\
& HotpotQA & \textbf{96.47\%} & 85.88\% & \textbf{8.92} & 17.14 \\
& MS-MARCO  & \textbf{97.33\%} & \textbf{97.33\%} & \textbf{9.00} & 16.99 \\
\midrule

\multirow{3}{*}{Phi-4-mini}
& NQ       & \textbf{98.97\%} & 91.75\% & \textbf{1.06} & 5.82 \\
& HotpotQA & \textbf{97.65\%} & 84.71\% & \textbf{1.01} & 6.43 \\
& MS-MARCO  & \textbf{100.00\%} & 93.33\% & \textbf{1.09} & 5.42 \\
\bottomrule
\end{tabular}
\end{adjustbox}
\caption{Comparison of \texttt{TRACE} and RAGForensics. Time denotes the average detection time per question, measured in seconds.}
\label{tab:detect_forensics}
\end{table}

We evaluate \texttt{TRACE} against RAGForensics~\cite{10.1145/3696410.3714756} using the configuration from Figure~\ref{fig_result1}, with the comparative summary detailed in Table~\ref{tab:detect_forensics}. RAGForensics utilizes specifically crafted prompts to query an LLM for detection. We re-implemented RAGForensics using the identical target LLMs as \texttt{TRACE}. The results demonstrate that while \texttt{TRACE} delivers superior detection performance across target models, its average execution time per question remains lower than that of RAGForensics. Crucially, while RAGForensics operates reactively, where users or systems must first doubt the RAG output and leverage the known target answer for backward corpus verification, \texttt{TRACE} requires no prior information regarding the target answer, uniquely identifying target answer tokens within the poisoned documents beforehand. Achieving superior efficacy with minimal time overhead and zero prior knowledge positions \texttt{TRACE} as an ideal candidate for real-time RAG poisoning defense.

\subsection{Evaluating Detection Accuracy with Phase 1's Output}

\begin{figure}[t]
    \centering    
    \includegraphics[width=0.48\textwidth]{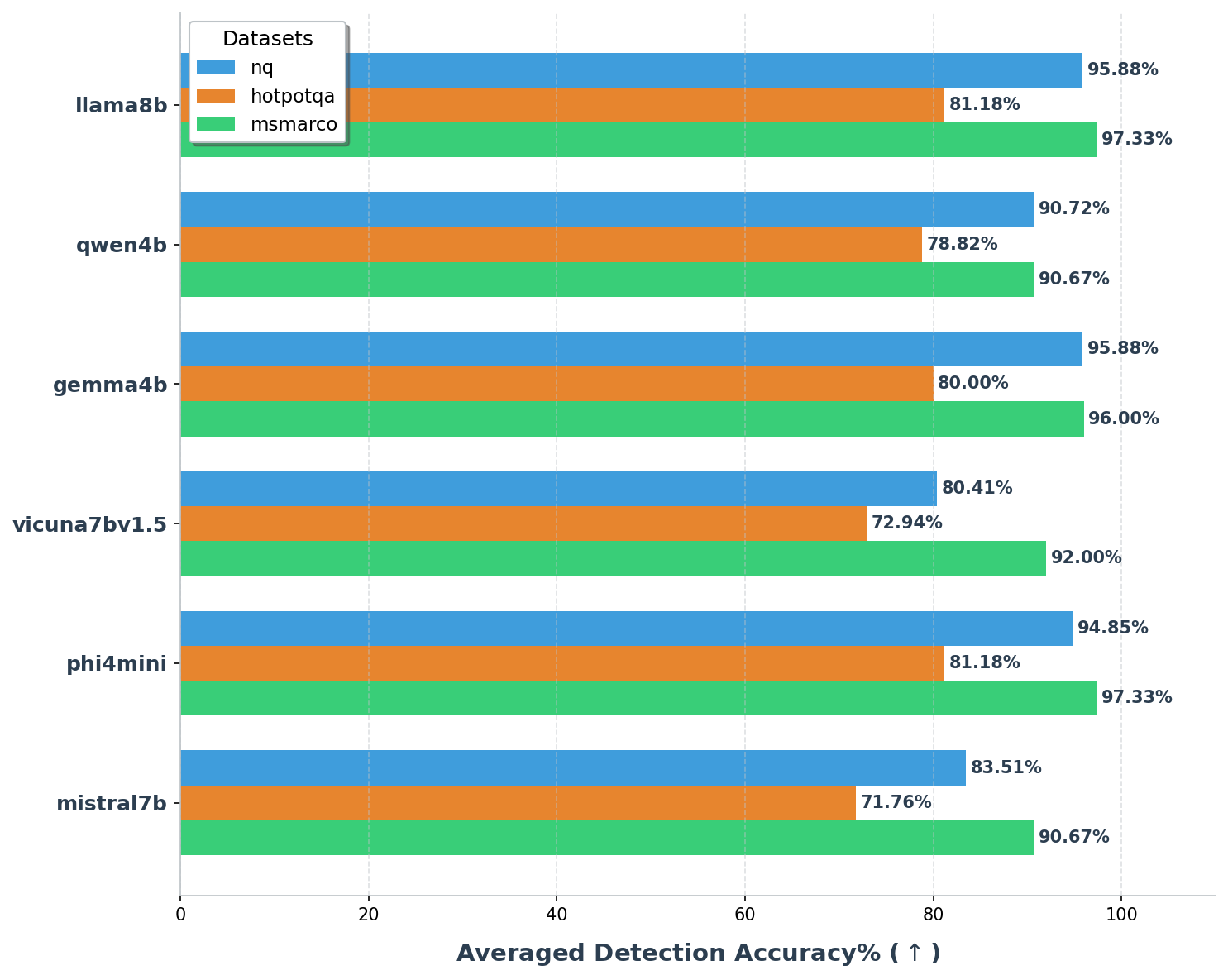}
    \caption{ACC results on Keyword Set $K$ with $a_1=3$, $a_2=2$, $k_1=5$, $k_2=3$.}
    \label{fig_result6}
\end{figure}

Utilizing the identical methodology established for evaluating ACC, we further examine whether the phrases within the keyword set $K$ also reside in the target answers. As illustrated in Figure~\ref{fig_result6}, the phrases in $K$ exhibit a higher ACC than those in Figure~\ref{fig_result5} across all models, which aligns perfectly with our filtering design. However, a noticeable discrepancy exists between the high TPR (over $90\%$) in Figure~\ref{fig_result1} and the ACC (between $70\%$ and $82\%$) on HotpotQA in both Figure~\ref{fig_result5} and \ref{fig_result6}. 

To demystify this phenomenon, we conducted an analysis on the keyword set $K$ and the target set $T$ generated under the HotpotQA setting. We observed that in a small subset of successfully flagged adversarial samples, neither $K$ nor $T$ contains the tokens of the actual target answer. Nonetheless, because other suspicious tokens were filtered into $T$, the corpus was still correctly classified as poisoned. We hypothesize that although these captured tokens do not manifest as the target answer, they elicit the maximum attention within the target LLMs, thereby acting as semantic anchors that steer the model generation towards the attacker's target answer. Universally, even in the worst-case under the PoisonedRAG attack, \texttt{TRACE} can still unmask over $70\%$ of the target answers.

\begin{table*}[t]
\centering
\small
\setlength{\tabcolsep}{3pt}
\renewcommand{\arraystretch}{1.05}
\begin{tabularx}{\textwidth}{
    @{}l*{5}{>{\centering\arraybackslash}X}@{}
}
\toprule
\textbf{Operating Point}
& \textbf{Final FPR}
& \shortstack{\textbf{Phase-1 Answer-}\\\textbf{Matching FPR}}
& \shortstack{\textbf{Final Answer-}\\\textbf{Matching FPR}}
& \shortstack{\textbf{Phase-2}\\\textbf{Reduction (pp)}}
& \shortstack{\textbf{Final FPs}\\\textbf{Matching Gold}} \\
\midrule
Balanced ($a_1=4,a_2=3$)
& 13.24\%
& 13.12\%
& 5.25\%
& 7.87
& 39.7\% \\
Low-FPR ($a_1=4,a_2=4$)
& 7.68\%
& 13.12\%
& 3.59\%
& 9.53
& 46.7\% \\
\bottomrule
\end{tabularx}
\caption{Benign answer-consensus analysis. Answer-matching FPR is measured over all benign queries, whereas the final column reports the proportion of final false positives whose detected target matches the gold answer.}
\label{tab:benign_consensus}
\end{table*}

\subsection{Analysis of Benign Answer Consensus}

We analyze whether \texttt{TRACE} may flag agreement among benign documents by identifying detected targets that match the gold answer under a normalized bidirectional substring criterion. As shown in Table~\ref{tab:benign_consensus}, Phase 2 substantially reduces answer-matching false positives under both operating points, although benign consensus remains a source of some final false positives. This finding further clarifies that \texttt{TRACE} screens for recurrent answer-driving influence but does not independently determine whether the consensus is benign or malicious.

\subsection{Benign-Agreement Hard Negatives}

\begin{table}[t]
\centering
\scriptsize
\setlength{\tabcolsep}{4pt}
\begin{tabular}{lccc}
\toprule
\textbf{Setting}
& \textbf{Queries}
& \textbf{FPR}
& \shortstack{\textbf{Gold-Matching}\\\textbf{Alert Rate}} \\
\midrule
Standard benign retrievals
& 257
& 13.16\%
& -- \\

Benign-agreement hard negatives
& 17
& 39.22\%
& 31.37\% \\
\bottomrule
\end{tabular}
\caption{Evaluation on benign-agreement hard negatives. The gold-matching alert rate is the proportion of all model-query evaluations in which \texttt{TRACE} raises an alert and identifies a final target keyword matching the gold answer.}
\label{tab:benign_agreement}
\end{table}

To evaluate \texttt{TRACE} under strong benign consensus, we constructed 17 hard-negative retrieval sets from HotpotQA and MS-MARCO. GPT-5 mini~\cite{openai2025gpt5} automatically identified clean passages that directly supported the gold answer based on the question, answer, title, and passage content. After normalized-text deduplication, each retained set contained at least three supporting passages; GPT-5 mini only annotated existing documents and did not generate or modify them.

Table~\ref{tab:benign_agreement} reports results across three target models under the balanced setting ($a_1=4,a_2=3$). The higher FPR on benign-agreement retrievals indicates that strong factual consensus is challenging for \texttt{TRACE}, although most evaluations remain unflagged. This suggests that alerts depend on the strength and recurrence of token-level influence rather than document agreement alone.

\subsection{Component Ablation Studies}

We ablate 3 components of \texttt{TRACE} using Llama-3.1-8B, Qwen-3.5-4B, and Phi-4-mini across the 3 datasets. All experiments use the balanced configuration ($a_1=4$, $a_2=3$, $k_1=5$, $k_2=3$), and the results are macro-averaged across the evaluated model-dataset combinations.

\paragraph{Affirmation String}
Table~\ref{tab:affirmation_ablation} shows that \texttt{TRACE} remains stable across different affirmation strings, with only minor variations in TPR, FPR, and Target ACC. This indicates limited sensitivity to the specific prompt formulation used for influence attribution.

\begin{table}[t]
\centering
\small
\setlength{\tabcolsep}{3pt}
\begin{tabularx}{\columnwidth}{@{}Xccc@{}}
\toprule
\textbf{Affirmation String} &
\textbf{TPR} &
\textbf{FPR} &
\textbf{Target ACC} \\
\midrule
\texttt{Yes, the answer is:}
& 88.13\% & 13.16\% & 80.97\% \\
\midrule
\texttt{The answer is:}
& 88.13\% & 13.64\% & 81.37\% \\
\midrule
\texttt{Answer:}
& 88.34\% & 13.46\% & 82.10\% \\
\midrule
\texttt{Based on the context, the answer is:}
& 87.47\% & 13.28\% & 80.74\% \\
\bottomrule
\end{tabularx}
\caption{Ablation of the affirmation string.}
\label{tab:affirmation_ablation}
\end{table}

\paragraph{Exclusion Lists}
As shown in Table~\ref{tab:exclusion_ablation}, removing the frequent-word exclusion has only a limited effect. In contrast, removing punctuation filtering substantially increases the FPR,
demonstrating its importance in preventing non-semantic tokens from being treated as recurrent influential evidence.

\begin{table}[t]
\centering
\small
\setlength{\tabcolsep}{3pt}
\begin{tabularx}{\columnwidth}{@{}Xccc@{}}
\toprule
\textbf{Exclusion Setting} &
\textbf{TPR} &
\textbf{FPR} &
\textbf{Target ACC} \\
\midrule
Full lists
& 88.13\% & 13.16\% & 80.97\% \\
Keep frequent words
& 87.31\% & 13.09\% & 79.68\% \\
Keep punctuations
& 98.94\% & 73.14\% & 76.52\% \\
Keep both
& 98.79\% & 72.58\% & 75.03\% \\
\bottomrule
\end{tabularx}
\caption{Ablation of the token-exclusion lists.}
\label{tab:exclusion_ablation}
\end{table}

\paragraph{Contiguous-Token Merging}
Table~\ref{tab:merging_ablation} shows only small performance changes
when contiguous-token merging is disabled. Thus, merging improves the
representation of multi-token target phrases but is not the primary
source of \texttt{TRACE}'s aggregate detection performance.

\begin{table}[!htbp]
\centering
\small
\setlength{\tabcolsep}{3pt}
\begin{tabularx}{\columnwidth}{@{}Xccc@{}}
\toprule
\textbf{Token Merging} &
\textbf{TPR} &
\textbf{FPR} &
\textbf{Target ACC} \\
\midrule
Enabled
& 88.13\% & 13.16\% & 80.97\% \\
Disabled
& 88.50\% & 13.87\% & 81.27\% \\
\bottomrule
\end{tabularx}
\caption{Ablation of contiguous-token merging.}
\label{tab:merging_ablation}
\end{table}

\section{Conclusion}
We introduced \texttt{TRACE}, a lightweight framework for detecting RAG poisoning attacks through token influence attribution. By identifying recurrent high-influence tokens and verifying their effect on model predictions, \texttt{TRACE} detects poisoned corpora without auxiliary models or external supervision. Experimental results show effective detection performance and demonstrate the ability to expose attacker-targeted answers across diverse LLMs.

\section*{Limitations}
The core mechanism of \texttt{TRACE} relies on the intuition that target LLMs exhibit an anomalous focus on the target answers designated by attackers within poisoned documents. Since this detection strategy is specifically designed for generative question answering settings, its underlying assumptions and empirical performance may differ from those observed in our experiments, which are conducted exclusively on question answering tasks, when applied to binary classification settings such as True or False question answering.

When poisoning attacks are rare, the base-rate effect causes false alerts to outnumber true detections unless the FPR is extremely low. Consequently, \texttt{TRACE} is better suited as a lightweight first-stage screening mechanism than as a standalone autonomous blocker. Its alerts should form a high-recall candidate set for subsequent verification or investigation.

The absolute recurrence thresholds $a_1$ and $a_2$ do not automatically scale with the number of retrieved documents. Applying fixed thresholds to larger retrieval sets can increase false positives by creating more opportunities for incidental token recurrence. Deployments with different retrieval depths therefore require depth-specific calibration or normalized recurrence thresholds.

\texttt{TRACE} is not a general-purpose RAG poisoning detector. It screens for coordinated cross-document answer-driving recurrence characteristic of PoisonedRAG-style corpus poisoning and does not independently determine whether the resulting consensus is true, false, benign, or malicious. Attacks that lack such recurrence, including isolated single-document poisoning and potentially paraphrased-target planting, fall outside its currently validated scope.

Because \texttt{TRACE} relies on the statistical recurrence of target answers, its sensitivity drops when attackers inject few poisoned documents. However, attackers cannot predict a user's retrieval size. Retrieving fewer poisoned documents inherently reduces the ASR. Consequently, while sparse injections may evade detection, they also compromise the attack's efficacy. This trade-off makes \texttt{TRACE} an effective first-line defense for RAG pipelines, efficiently screening out high-impact poisoning attempts early.

\section*{Ethical Considerations}
This work aims to improve the security and trustworthiness of RAG systems by detecting corpus poisoning attacks. Our method is evaluated using publicly available question-answering datasets and previously published attack frameworks. \texttt{TRACE} does not generate harmful content or introduce new attack mechanisms; instead, it identifies suspicious retrieved documents and reveals attacker-targeted answers to support defensive analysis.

We acknowledge that token influence attribution could potentially be used to study how poisoned documents affect model behavior. However, the techniques presented in this paper are intended solely for security evaluation and defense. We do not release any new poisoning datasets, attack tools, or infrastructure that would substantially lower the barrier for conducting attacks. We believe the benefits of improving RAG security, reliability, and transparency outweigh the potential risks associated with this research.

\section*{Acknowledgments}

This research was partially supported by Hon Hai Research Institute, NSTC 114-2222-E-A49-011-MY3, TACC, NSTC 114-2634-F-A49-006, and NSTC 115-2923-E-A49-010-MY5.

\bibliography{anthology,custom}
\bibliographystyle{acl_natbib}

\appendix
\section{Pseudocode}\label{sec:pseudocode}
Here, we include the pseudocode for the Phase 1 (extract possible keywords) and Phase 2 (secondary verification as Algorithm~\ref{alg_1} and Algorithm ~\ref{alg_2}, respectively.

\begin{algorithm}[t]
\caption{Phase 1: Extract Possible Keywords}
\label{alg_1}
\linespread{1.2}
\footnotesize
\textbf{Input:} Retrieved document set $D$, target LLM parameters $\theta$, user question $q$, affirmation string $y$, punctuation set $P$, frequent words set $F$, minimum occurrence threshold $a_1$, number of top-influence tokens $k_1$. \\
\textbf{Output:} Keyword set $K$
\begin{algorithmic}[1]
\STATE $K_{temp} \leftarrow \emptyset$ 
\FOR{\textbf{each} $d \in D$}
    \STATE $n \leftarrow \text{length}(d)$
    \STATE $x_{1:n} \leftarrow embed_{\theta}(d_{1:n})$
    \STATE $V \leftarrow \emptyset$ 
    \FOR{$i = 1$ \TO $n$}
        \STATE $\texttt{Influence}(i) \leftarrow \left\| \nabla_{x_i} \log P_{\theta}(y | x_{1:n}) \right\|_2$
        \IF{$d_i \notin q$ \AND $d_i \notin P$ \AND $d_i \notin F$}
            \STATE $V \leftarrow V \cup \{i\}$
        \ENDIF
    \ENDFOR
    
    \STATE $I_{top} \leftarrow \underset{i \in V}{\operatorname{\texttt{argtop-}}k_1} \, (\texttt{Influence}(x_i))$
    \STATE $K_d \leftarrow \emptyset$
    \STATE Partition $I_{top}$ to a set of contiguous index segments $S$
    \FOR{\textbf{each} segment $(i, i+1, \dots, j) \in S$}
        \STATE $e \leftarrow \text{concatenate}(d_{i:j})$
        \STATE $K_d \leftarrow K_d \cup \{e\}$
    \ENDFOR
    \STATE Append $K_d$ to $K_{temp}$
\ENDFOR

\STATE $K \leftarrow \emptyset$
\STATE $U \leftarrow$  the set of all unique candidate in $K_{temp}$
\FOR{\textbf{each} $e \in U$}
    \STATE $count \leftarrow |\{ d \in D \mid e \in d \}|$ 
    \IF{$count \ge a_1$}
        \STATE $K \leftarrow K \cup \{e\}$
    \ENDIF
\ENDFOR
\RETURN $K$
\end{algorithmic}
\end{algorithm}

\begin{algorithm}[t]
\caption{Phase 2: Secondary Verification}
\label{alg_2}
\linespread{1.2}
\footnotesize
\textbf{Input:} Retrieved document set $D$, target LLM parameterized by $\theta$, candidate keyword set $K$, user question $q$, punctuation set $P$, frequent words set $F$, minimum occurrence threshold $a_2$, number of top-influence tokens $k_2$. \\
\mbox{\textbf{Output:} Target keyword set $T$, detection flag $f$}
\begin{algorithmic}[1]
\STATE $T \leftarrow \emptyset$
\FOR{\textbf{each} $p \in K$}
    \STATE $K_{temp} \leftarrow \emptyset$
    \FOR{\textbf{each} $d \in D$}
        \STATE $n \leftarrow \text{length}(d)$
        \STATE $x_{1:n} \leftarrow embed_{\theta}(d_{1:n})$
        \STATE $V \leftarrow \emptyset$
        \FOR{$i = 1$ \TO $n$}
            \STATE $Influence(i) \leftarrow \left\| \nabla_{x_i} \log P_{\theta}(p \mid x_{1:n}) \right\|_2$
            \IF{$d_i \notin q$ \AND $d_i \notin P$ \AND $d_i \notin F$}
                \STATE $V \leftarrow V \cup \{i\}$
            \ENDIF
        \ENDFOR
        
        \STATE $I_{top} \leftarrow \underset{i \in V}{\operatorname{\texttt{argtop-}}k_2} \, (\texttt{Influence}(x_i))$
        \STATE $K_d \leftarrow \{d_i \mid i \in I_{top}\ \land d_i \in p \}$ 
        \STATE Append $K_d$ to $K_{temp}$
    \ENDFOR
    
    \STATE $U \leftarrow \text{the set of all unique tokens in } K_{temp}$
    \FOR{\textbf{each} $e \in U$}
        \STATE $count \leftarrow |\{ K_d \in K_{temp} \mid e \in K_d \}|$
        \IF{$count \ge a_2$}
            \STATE $T \leftarrow T \cup \{e\}$
        \ENDIF
    \ENDFOR
\ENDFOR

\IF{$T \neq \emptyset$}
    \STATE $f \leftarrow \texttt{TRUE}$
\ELSE
    \STATE $f \leftarrow \texttt{FALSE}$
\ENDIF

\RETURN $T, f$
\end{algorithmic}
\end{algorithm}

\section{Punctuations and Frequent Words}
\label{sec:appendix_A}

\subsection{Punctuations}
\begin{mypromptbox}
\texttt{\textcolor{red}{!}},
\texttt{\textcolor{red}{"}},
\texttt{\textcolor{red}{\#}},
\texttt{\textcolor{red}{\%}},
\texttt{\textcolor{red}{\&}},
\texttt{\textcolor{red}{'}},
\texttt{\textcolor{red}{(}},
\texttt{\textcolor{red}{)}},
\texttt{\textcolor{red}{*}},
\texttt{\textcolor{red}{,}},
\texttt{\textcolor{red}{-}},
\texttt{\textcolor{red}{.}},
\texttt{\textcolor{red}{/}},
\texttt{\textcolor{red}{:}},
\texttt{\textcolor{red}{;}},
\texttt{\textcolor{red}{?}},
\texttt{\textcolor{red}{@}},
\texttt{\textcolor{red}{[}},
\texttt{\textcolor{red}{\textbackslash{}}},
\texttt{\textcolor{red}{]}},
\texttt{\textcolor{red}{\_}},
\texttt{\textcolor{red}{\{}},
\texttt{\textcolor{red}{\}}}
\end{mypromptbox}

\subsection{Frequent Words}
\begin{mypromptbox}
"a", "an", "the",
"in", "on", "at", "of", "to", "from",
"for", "with", "by", "about",
"and", "or", "but",
"is", "are", "was", "were", "be",
"he", "she", "it", "they", "we",
"you", "your", "I", "me", "my",
"this", "that", "these", "those",
"what", "which", "who", "whom", "whose",
"how", "when", "where", "why"
\end{mypromptbox}

\section{Models Used in Experiments}
\label{sec:appendix_B}

\textbf{\texttt{Llama-3.1-8B}} \\\texttt{meta-llama/Meta-Llama-3.1-8B-Instruct}\footnote{\href{https://huggingface.co/meta-llama/Llama-3.1-8B-Instruct}{https://huggingface.co/meta-llama/Llama-3.1-8B-Instruct}}\\
\textbf{\texttt{Gemma-3-4b}} \\\texttt{google/gemma-3-4b-it}\footnote{\href{https://huggingface.co/google/gemma-3-4b-it}{https://huggingface.co/google/gemma-3-4b-it}}\\
\textbf{\texttt{Qwen-3.5-4B}} \\\texttt{Qwen/Qwen3.5-4B}\footnote{\href{https://huggingface.co/Qwen/Qwen3.5-4B}{https://huggingface.co/Qwen/Qwen3.5-4B}}\\
\textbf{\texttt{Vicuna-7b}} \\\texttt{lmsys/vicuna-7b-v1.5}\footnote{\href{https://huggingface.co/lmsys/vicuna-7b-v1.5}{https://huggingface.co/lmsys/vicuna-7b-v1.5}}\\
\textbf{\texttt{Mistral-7B}} \\\texttt{mistralai/Mistral-7B-Instruct-v0.3}\footnote{\href{https://huggingface.co/mistralai/Mistral-7B-Instruct-v0.3}{https://huggingface.co/mistralai/Mistral-7B-Instruct-v0.3}}\\
\textbf{\texttt{Phi-4-mini}} \\\texttt{microsoft/Phi-4-mini-instruct}\footnote{\href{https://huggingface.co/microsoft/Phi-4-mini-instruct}{https://huggingface.co/microsoft/Phi-4-mini-instruct}}

\section{Effects of Retrieval Depth and Poisoning Dilution}

\subsection{More Retrieved Documents at a Fixed Poisoning Rate}

We evaluated retrieval depths of 5, 10, and 20 using the three LLMs (Llama, Qwen, and Phi) and three datasets. Each attacked query contains the original 5 poisoned documents; top-10 and top-20 additionally contain 5 and 15 naturally retrieved documents, respectively.

\begin{table}[t]
\centering
\scriptsize
\begin{tabular}{cccc}
\toprule
Retrieved Documents & TPR & FPR & Latency per Query (s) \\
\midrule
5  & 88.13\% & 13.16\% & 3.76 \\
10 & 99.25\% & 58.32\% & 10.09 \\
20 & 99.36\% & 84.83\% & 50.00 \\
\bottomrule
\end{tabular}
\caption{Results under more retrieved documents, with $a_1=4$, $a_2=3$, $k_1=5$, and $k_2=3$.}
\label{tab:more_retrieved_documents}
\end{table}

The results are shown in Table~\ref{tab:more_retrieved_documents}. Because the number of poisoned documents is fixed at five, increasing the retrieval depth does not make the attack stronger. Instead, the additional benign documents introduce competing evidence that may dilute or override the poisoned content, potentially reducing downstream attack success; therefore, the higher TPR should be interpreted as improved detector-level sensitivity rather than increased attack effectiveness. The increase in FPR likely results from more benign documents creating additional opportunities for accidental token overlap with the target answer, thereby triggering the fixed recurrence thresholds. Since practical RAG systems typically use a small fixed top-$k$, \texttt{TRACE} operates on the retrieved document set rather than the total number of poisoned documents injected into the corpus. Thus, under a fixed retrieval depth, adding more poisoned documents elsewhere in the corpus does not necessarily change \texttt{TRACE}’s input if the same number of poisoned documents is retrieved.

\subsection{Fewer Poisoned Documents at a Fixed Number of Retrieved Documents}

\begin{table}[t]
\scriptsize
\centering
\begin{tabular}{@{}cccc@{}}
\toprule
Thresholds ($a_1$, $a_2$) & Poisoned / Benign & TPR & Target Acc. \\
\midrule
$(4,3)$ & $1$ / $4$ &  9.98\% &  0.38\% \\
$(4,3)$ & $3$ / $2$ & 23.01\% &  8.60\% \\
$(2,2)$ & $1$ / $4$ & 68.40\% &  7.57\% \\
$(3,2)$ & $3$ / $2$ & 90.15\% & 73.29\% \\
\bottomrule
\end{tabular}
\caption{Results under fewer poisoned documents, with $k_1=5$, and $k_2=3$.}
\label{tab:fewer_poisoned_documents}
\end{table}

To evaluate \texttt{TRACE} under weaker poisoning conditions, we conduct a poisoned-document dilution experiment with a fixed retrieval depth of five. As shown in Table~\ref{tab:fewer_poisoned_documents}, reducing the number of poisoned documents degrades detection and target identification, while relaxing the recurrence thresholds improves sensitivity, particularly when three poisoned documents are retrieved. However, dilution also weakens the underlying attack signal and reduces opportunities for cross-document recurrence. The observed degradation therefore reflects both reduced attack effectiveness and lower detector sensitivity. These results suggest that \texttt{TRACE} is most effective against coordinated attacks in which the same target recurs across multiple retrieved documents. Detecting single-document poisoning and paraphrased-target attacks remains an important direction for future work.

\section{Results under Different Hyperparameter Configurations}
\label{sec:appendix_C}
\subsection{TPR}

\begin{figure}[H]
    \centering    
    \includegraphics[width=0.48\textwidth]{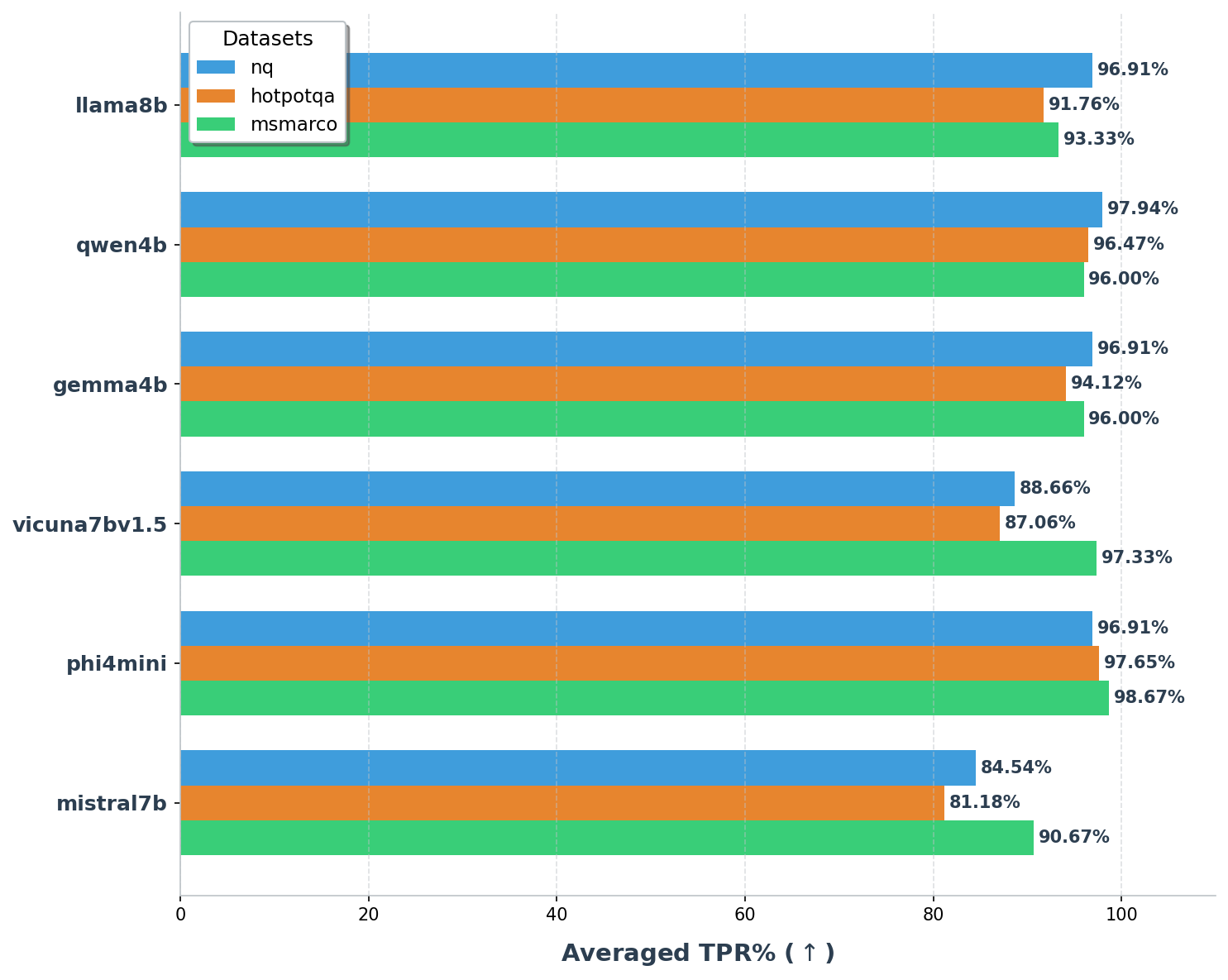}
    \caption{TPR results with $3$ retrieved documents, $a_1=2$, $a_2=2$, $k_1=5$, $k_2=3$.}
    \label{appendix_result1}
\end{figure}

\begin{figure}[H]
    \centering    
    \includegraphics[width=0.48\textwidth]{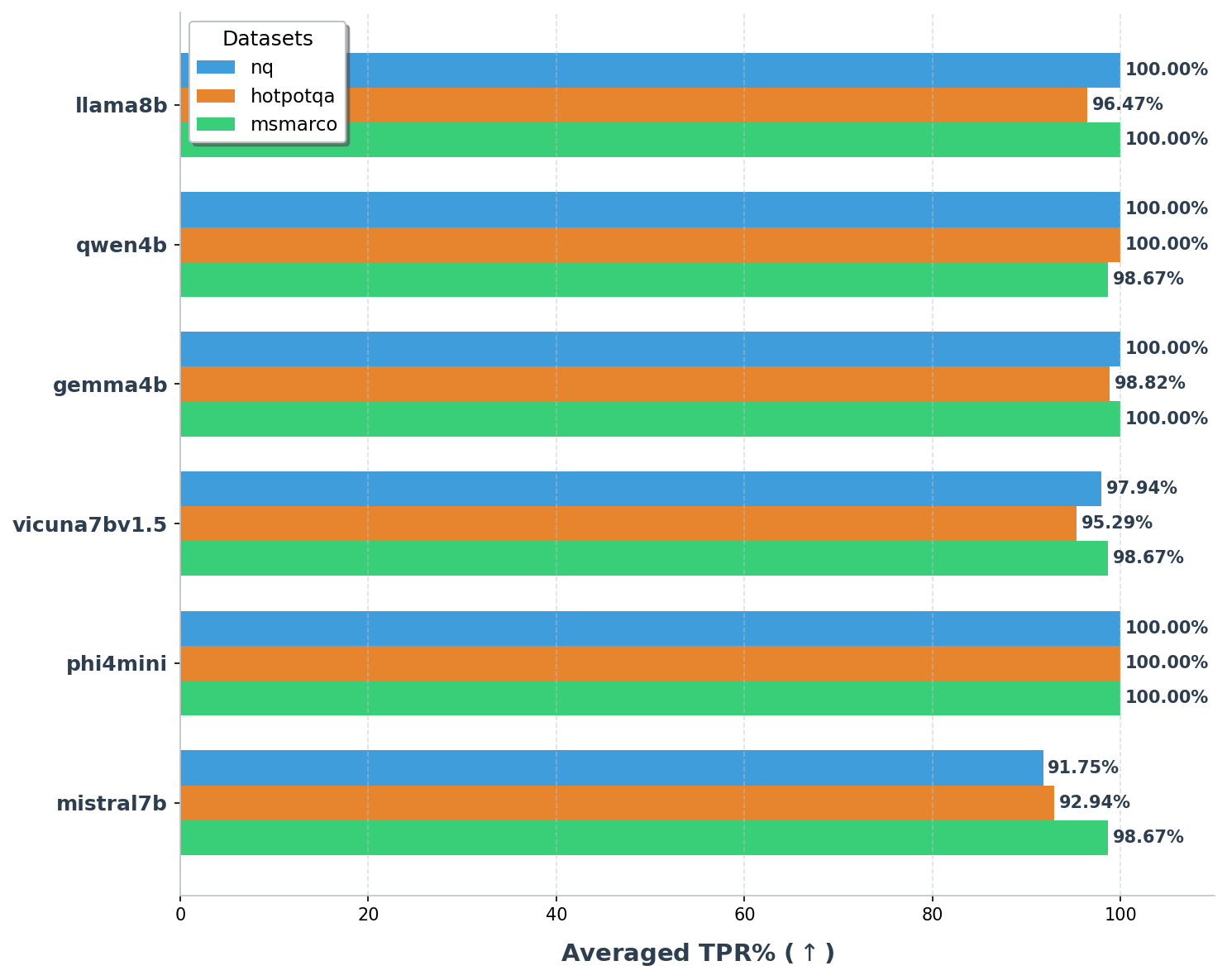}
    \caption{TPR results with $a_1=2$, $a_2=2$, $k_1=5$, $k_2=3$.}
    \label{appendix_result2}
\end{figure}

\begin{figure}[H]
    \centering    
    \includegraphics[width=0.48\textwidth]{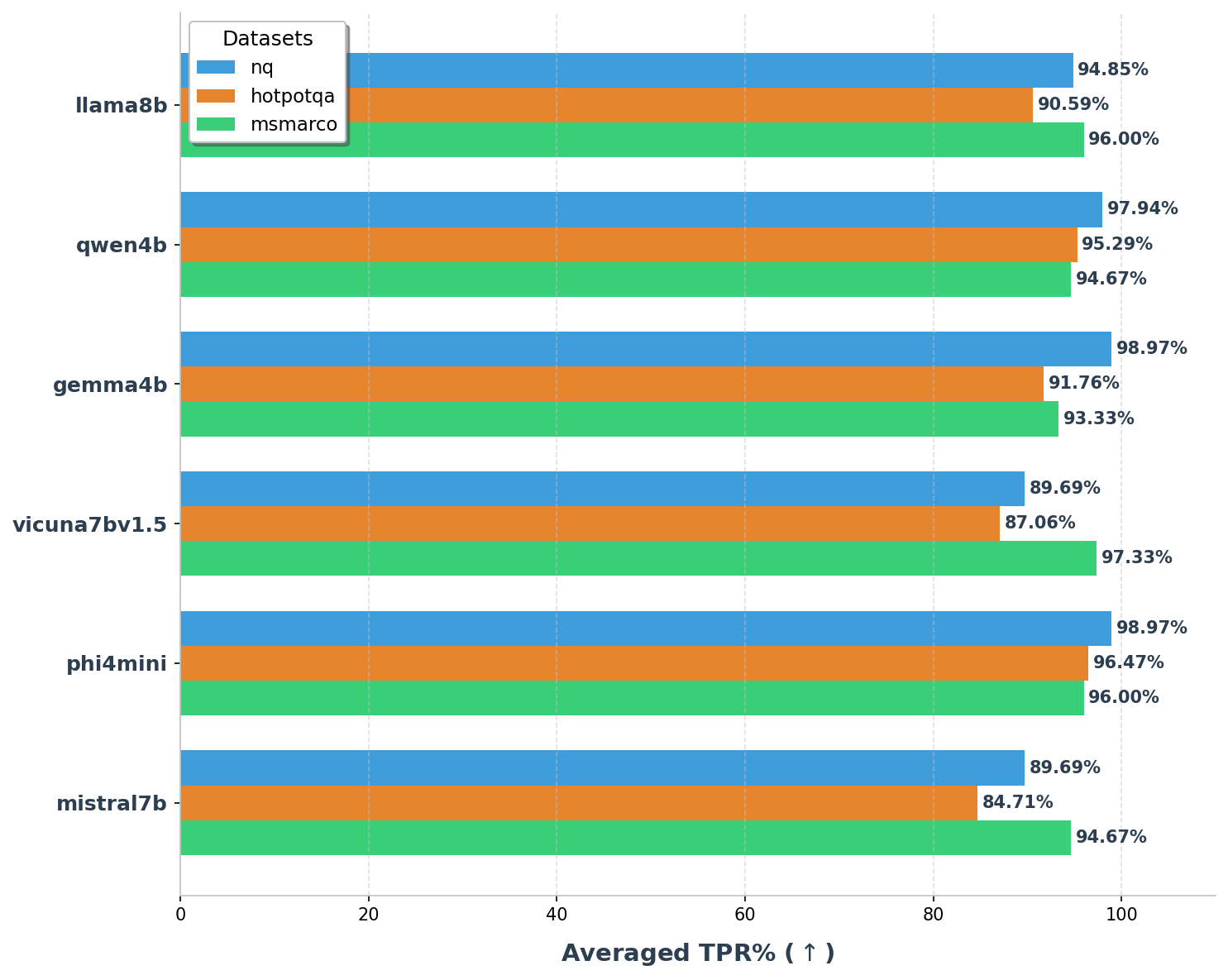}
    \caption{TPR results with $a_1=2$, $a_2=3$, $k_1=5$, $k_2=3$.}
    \label{appendix_result3}
\end{figure}

\begin{figure}[H]
    \centering    
    \includegraphics[width=0.48\textwidth]{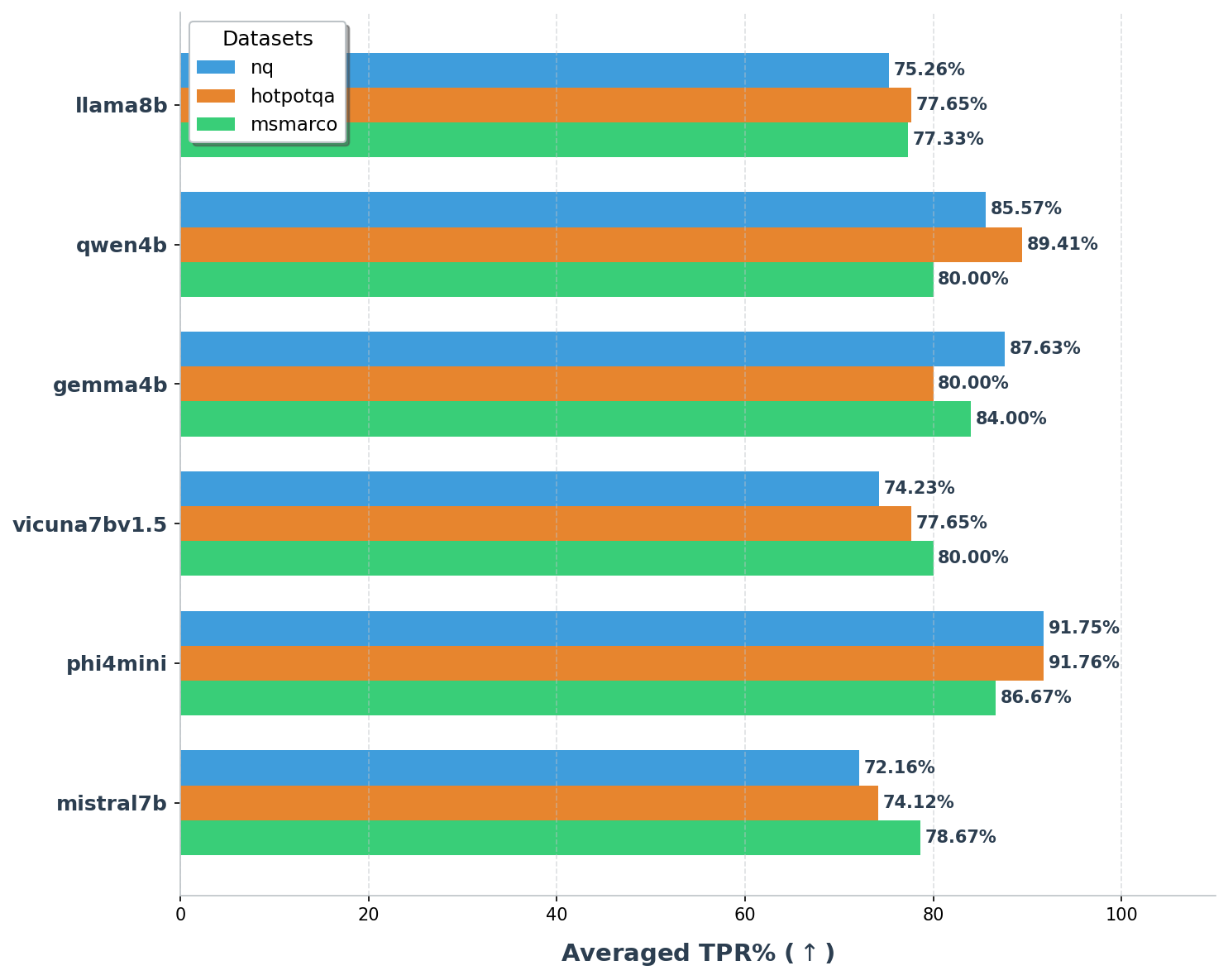}
    \caption{TPR results with $a_1=2$, $a_2=4$, $k_1=5$, $k_2=3$.}
    \label{appendix_result4}
\end{figure}

\begin{figure}[H]
    \centering    
    \includegraphics[width=0.48\textwidth]{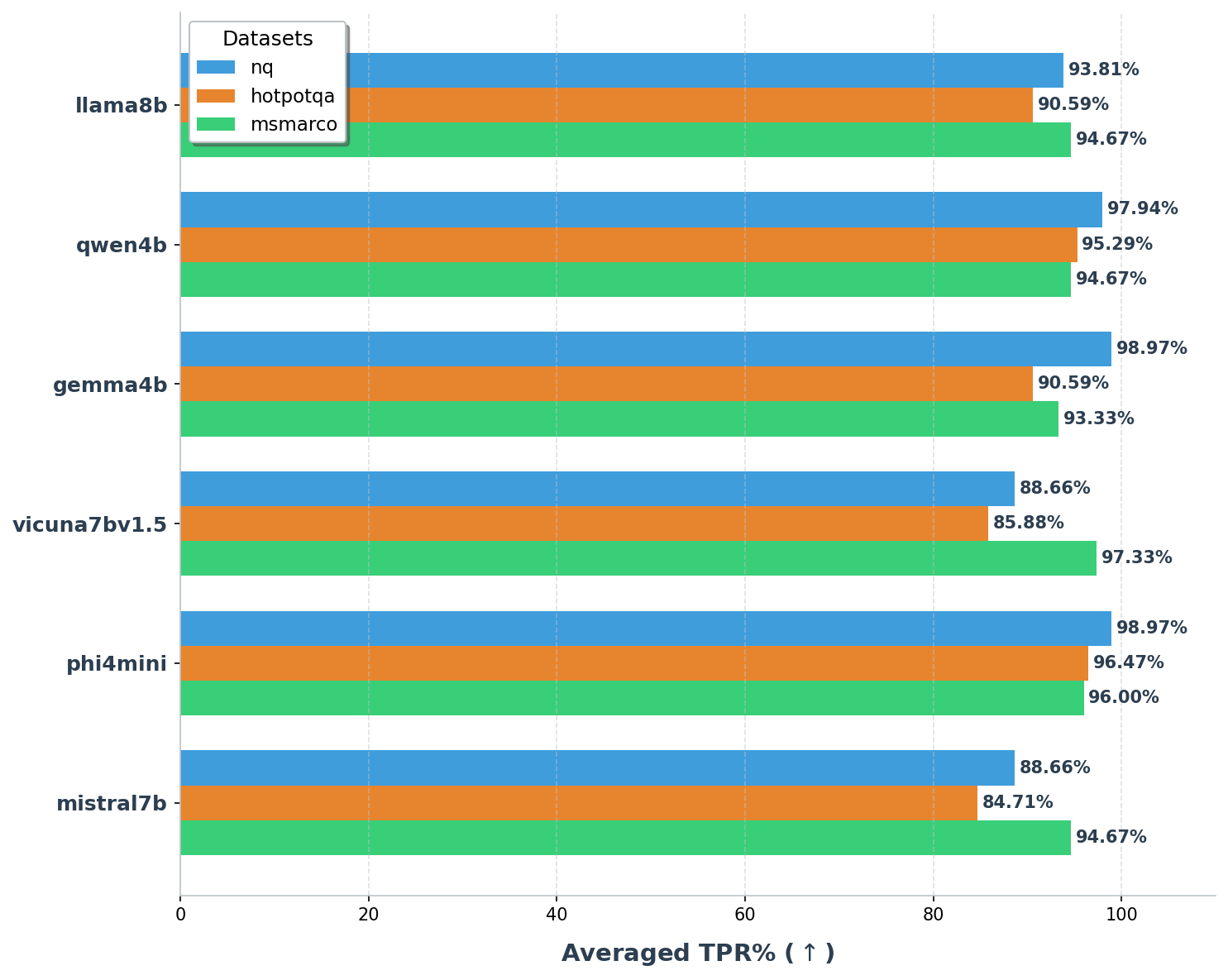}
    \caption{TPR results with $a_1=3$, $a_2=3$, $k_1=5$, $k_2=3$.}
    \label{appendix_result5}
\end{figure}

\begin{figure}[H]
    \centering    
    \includegraphics[width=0.48\textwidth]{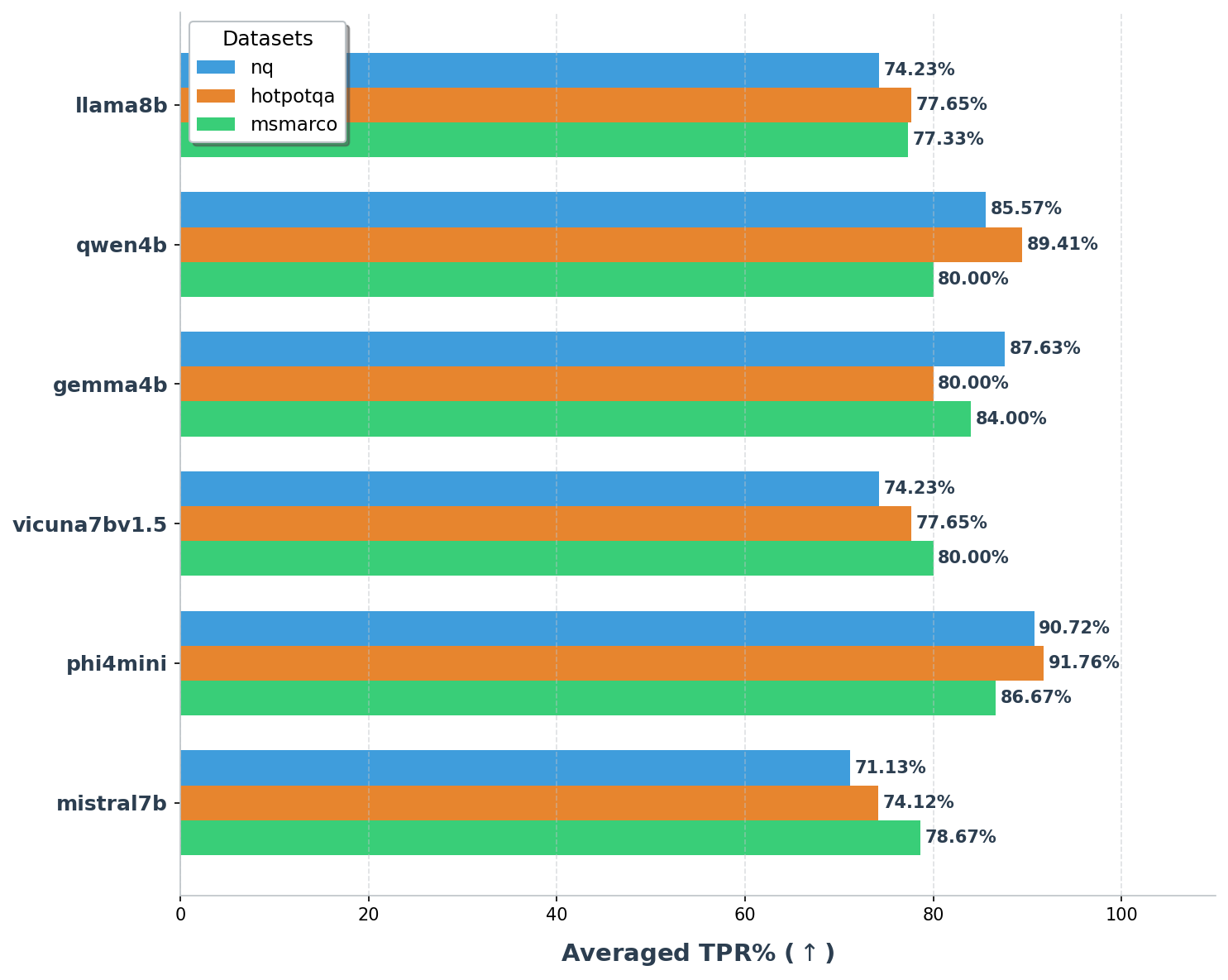}
    \caption{TPR results with $a_1=3$, $a_2=4$, $k_1=5$, $k_2=3$.}
    \label{appendix_result6}
\end{figure}

\begin{figure}[H]
    \centering    
    \includegraphics[width=0.48\textwidth]{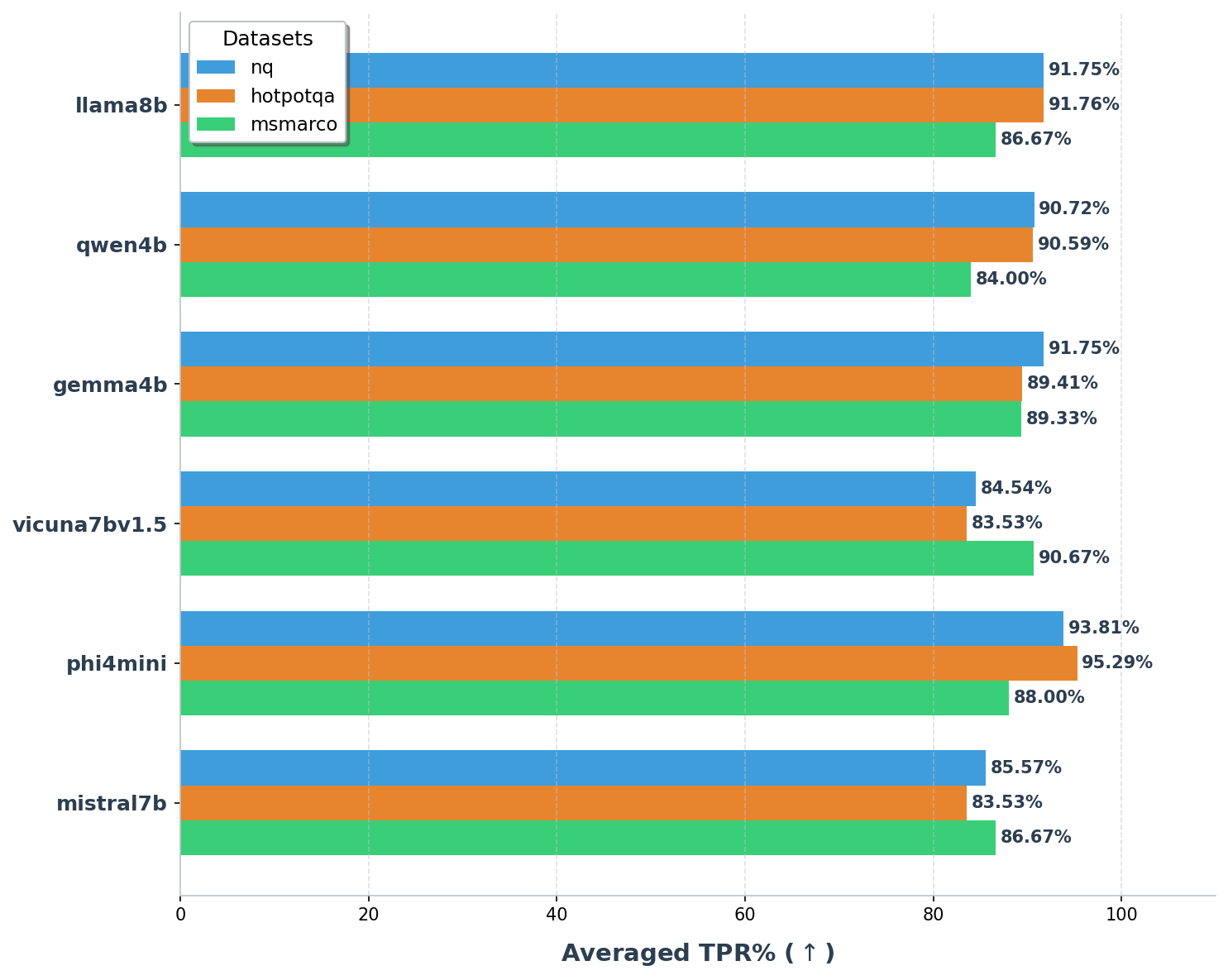}
    \caption{TPR results with $a_1=4$, $a_2=2$, $k_1=5$, $k_2=3$.}
    \label{appendix_result7}
\end{figure}

\begin{figure}[H]
    \centering    
    \includegraphics[width=0.48\textwidth]{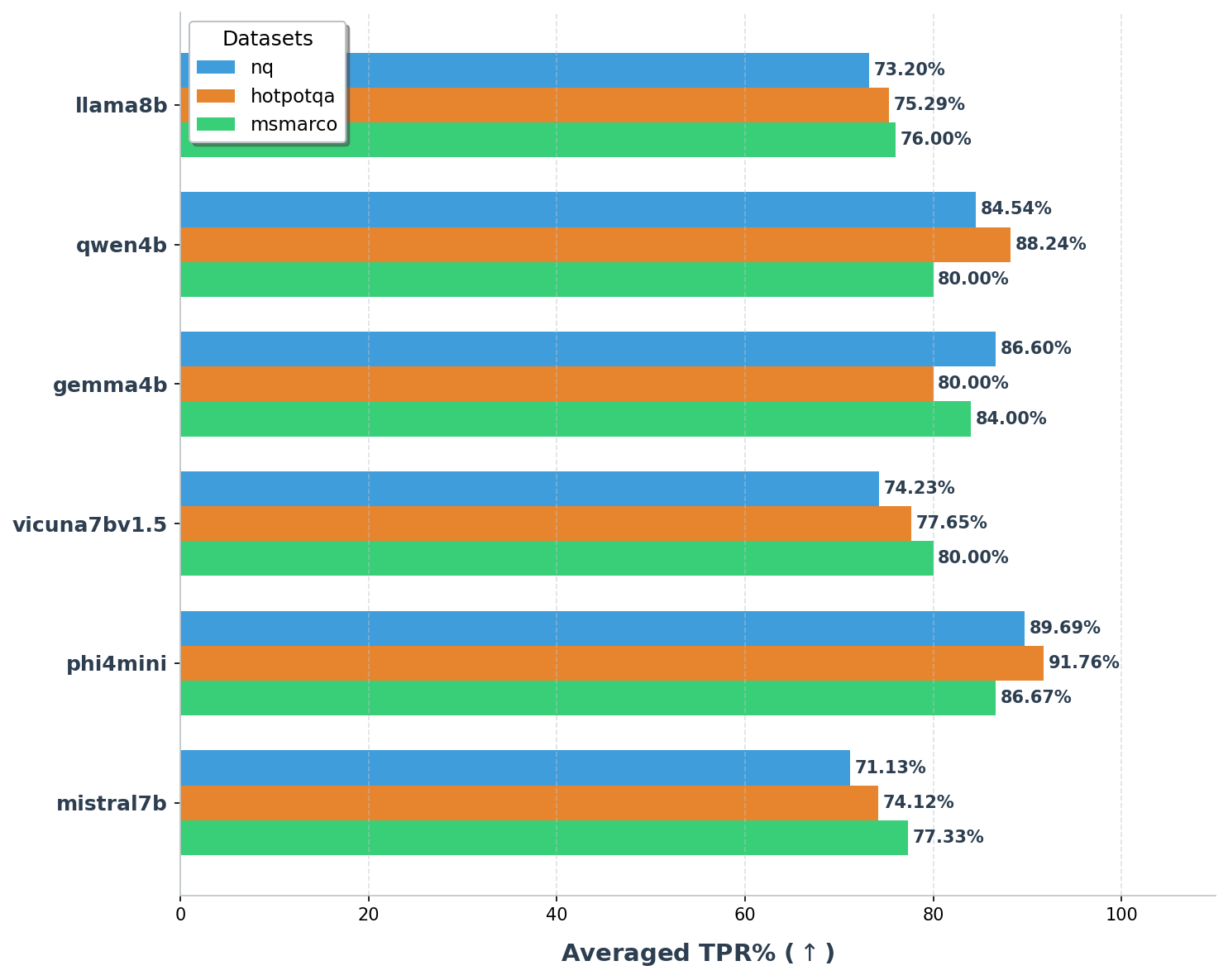}
    \caption{TPR results with $a_1=4$, $a_2=4$, $k_1=5$, $k_2=3$.}
    \label{appendix_result8}
\end{figure}

\subsection{FPR}

\begin{figure}[H]
    \centering    
    \includegraphics[width=0.48\textwidth]{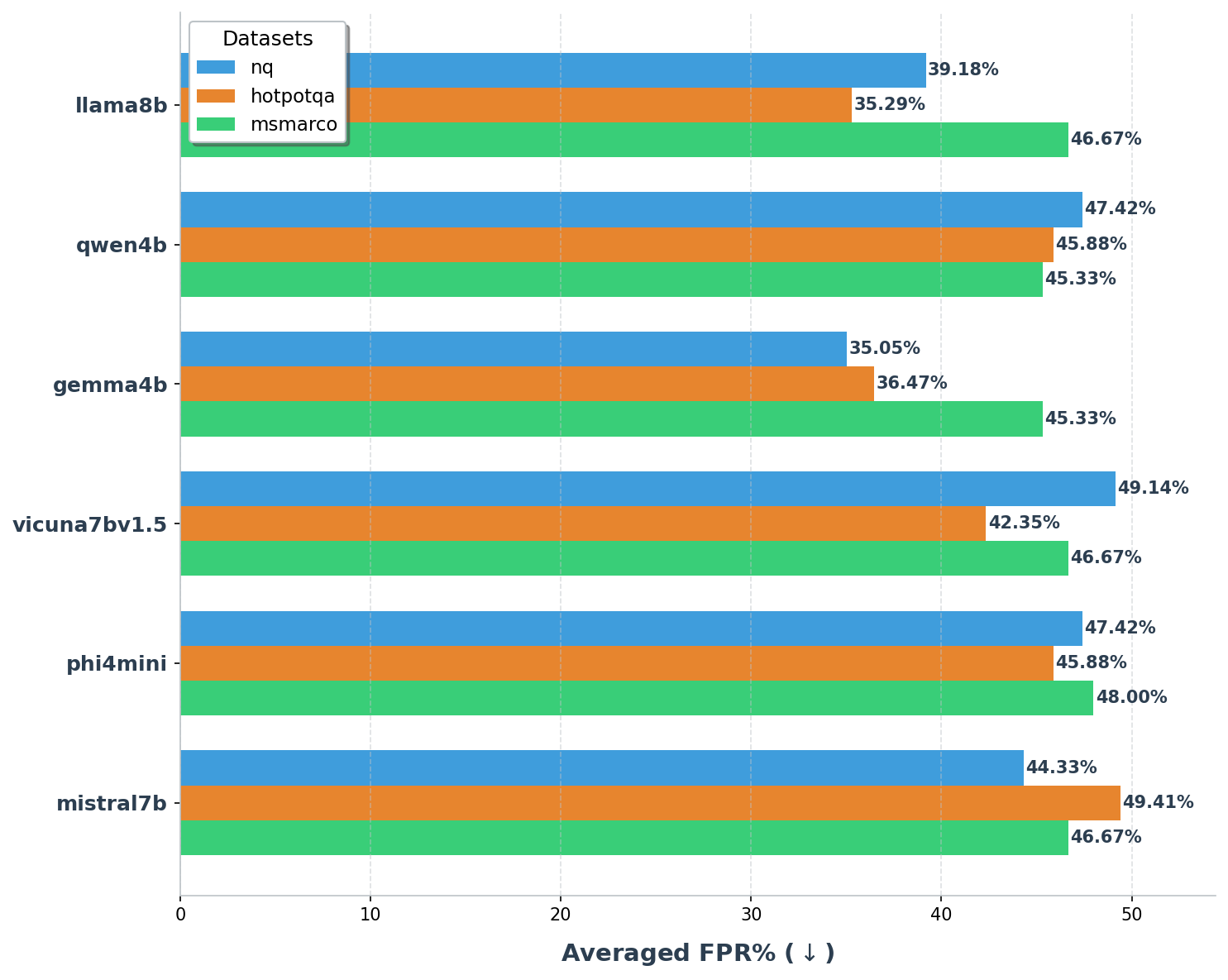}
    \caption{FPR results with $3$ retrieved documents, $a_1=2$, $a_2=2$, $k_1=5$, $k_2=3$.}
    \label{appendix_result9}
\end{figure}

\begin{figure}[H]
    \centering    
    \includegraphics[width=0.48\textwidth]{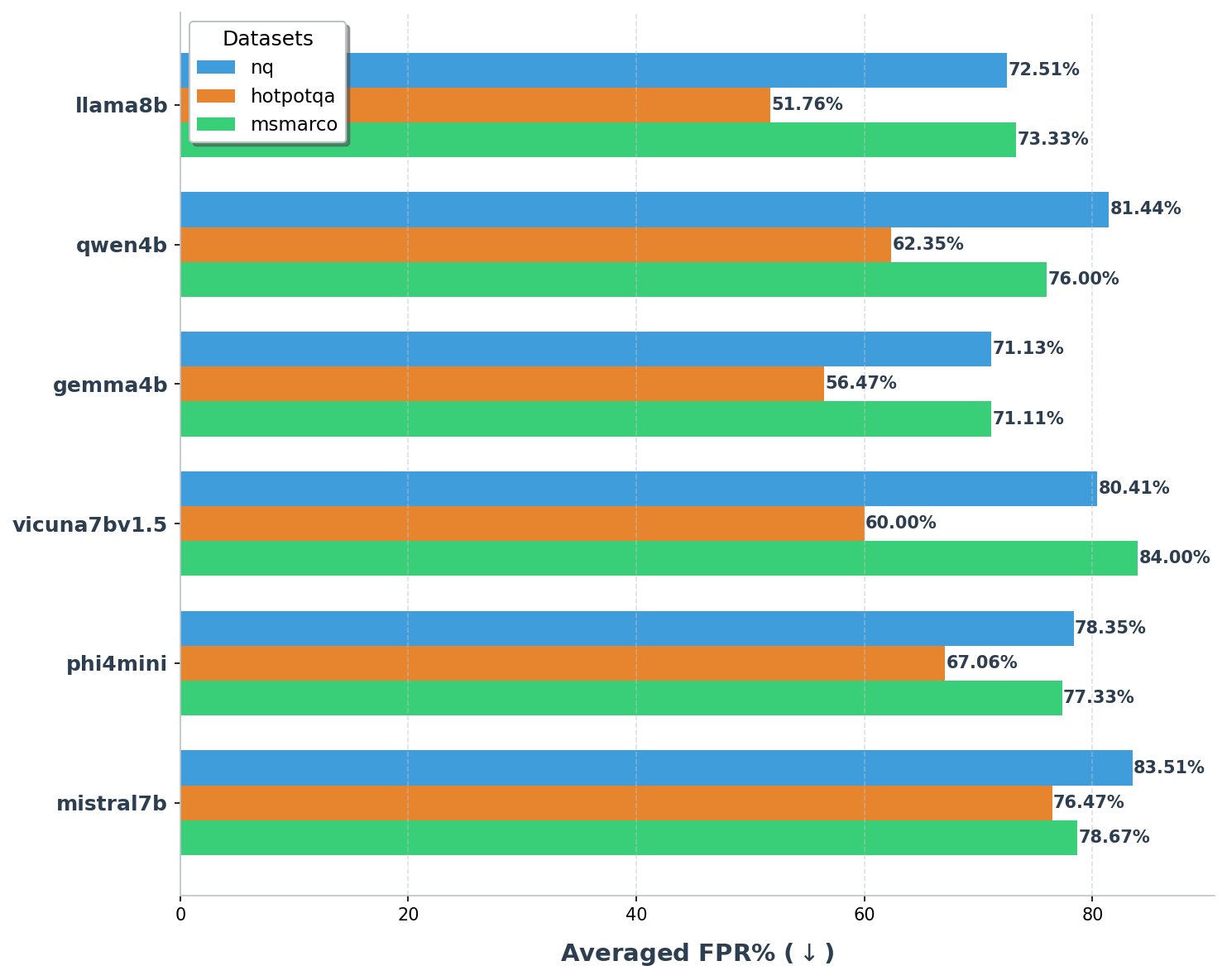}
    \caption{FPR results with $a_1=2$, $a_2=2$, $k_1=5$, $k_2=3$.}
    \label{appendix_result10}
\end{figure}

\begin{figure}[H]
    \centering    
    \includegraphics[width=0.48\textwidth]{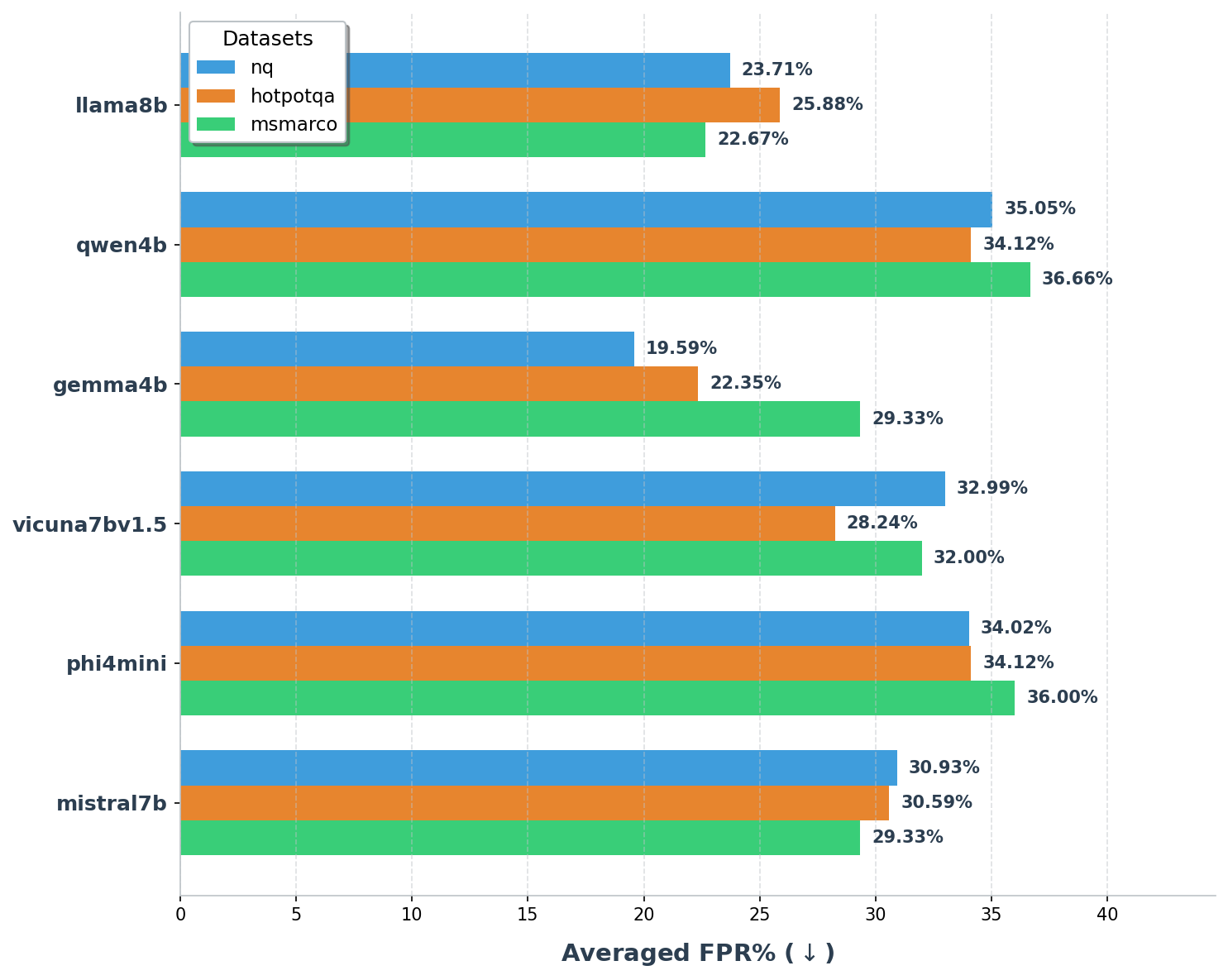}
    \caption{FPR results with $a_1=2$, $a_2=3$, $k_1=5$, $k_2=3$.}
    \label{appendix_result11}
\end{figure}

\begin{figure}[H]
    \centering    
    \includegraphics[width=0.48\textwidth]{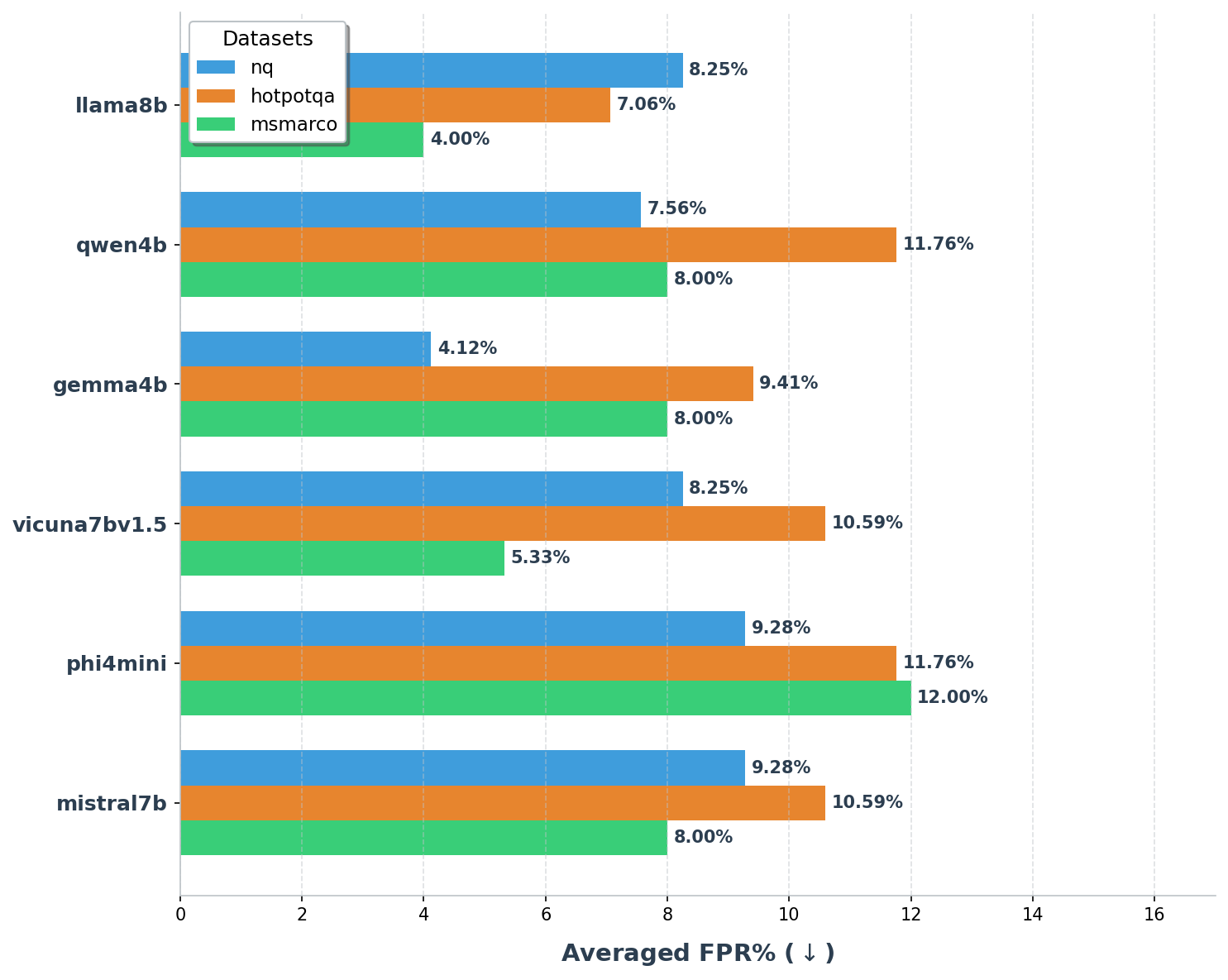}
    \caption{FPR results with $a_1=2$, $a_2=4$, $k_1=5$, $k_2=3$.}
    \label{appendix_result12}
\end{figure}

\begin{figure}[H]
    \centering    
    \includegraphics[width=0.48\textwidth]{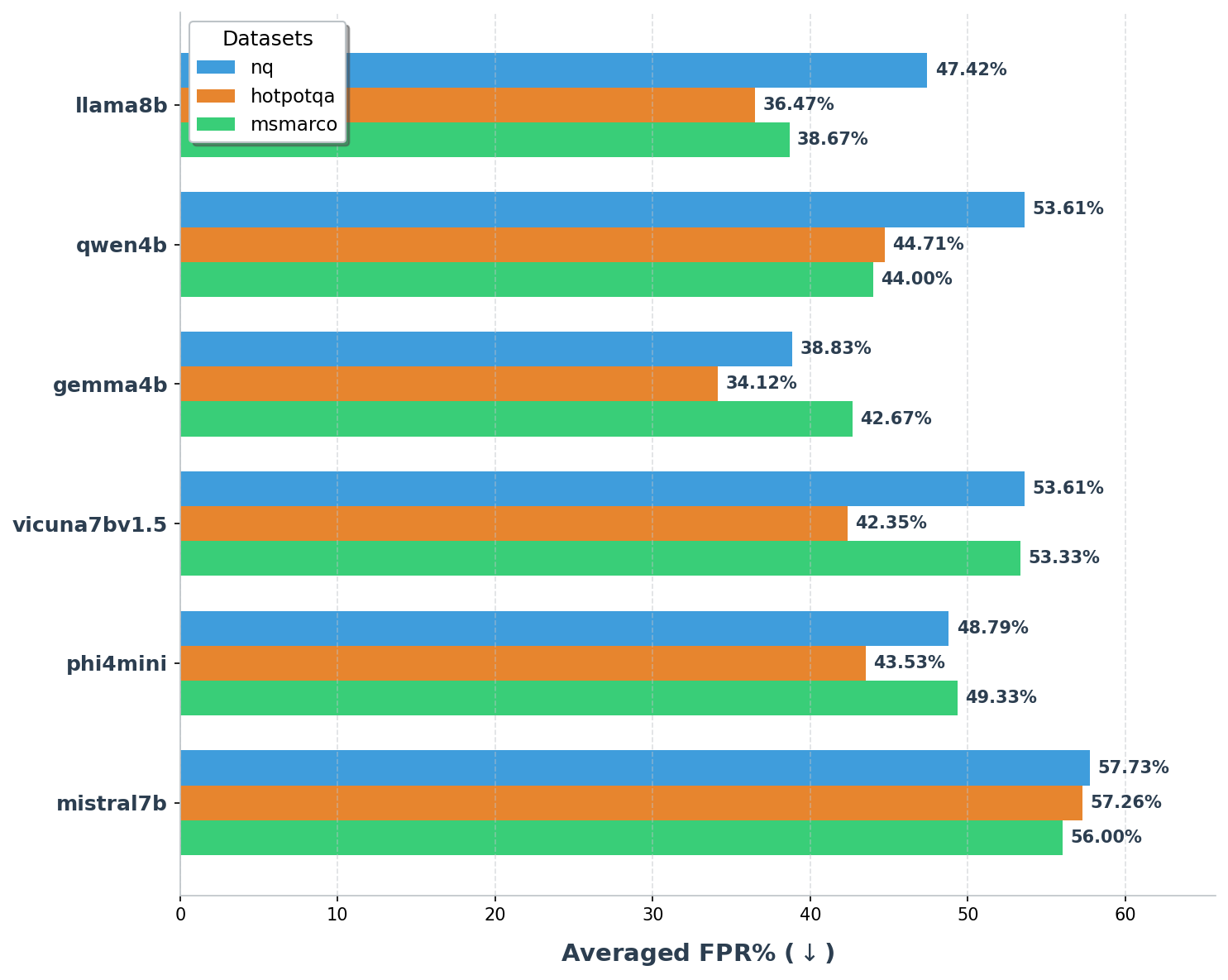}
    \caption{FPR results with $a_1=3$, $a_2=2$, $k_1=5$, $k_2=3$.}
    \label{appendix_result13}
\end{figure}

\begin{figure}[H]
    \centering    
    \includegraphics[width=0.48\textwidth]{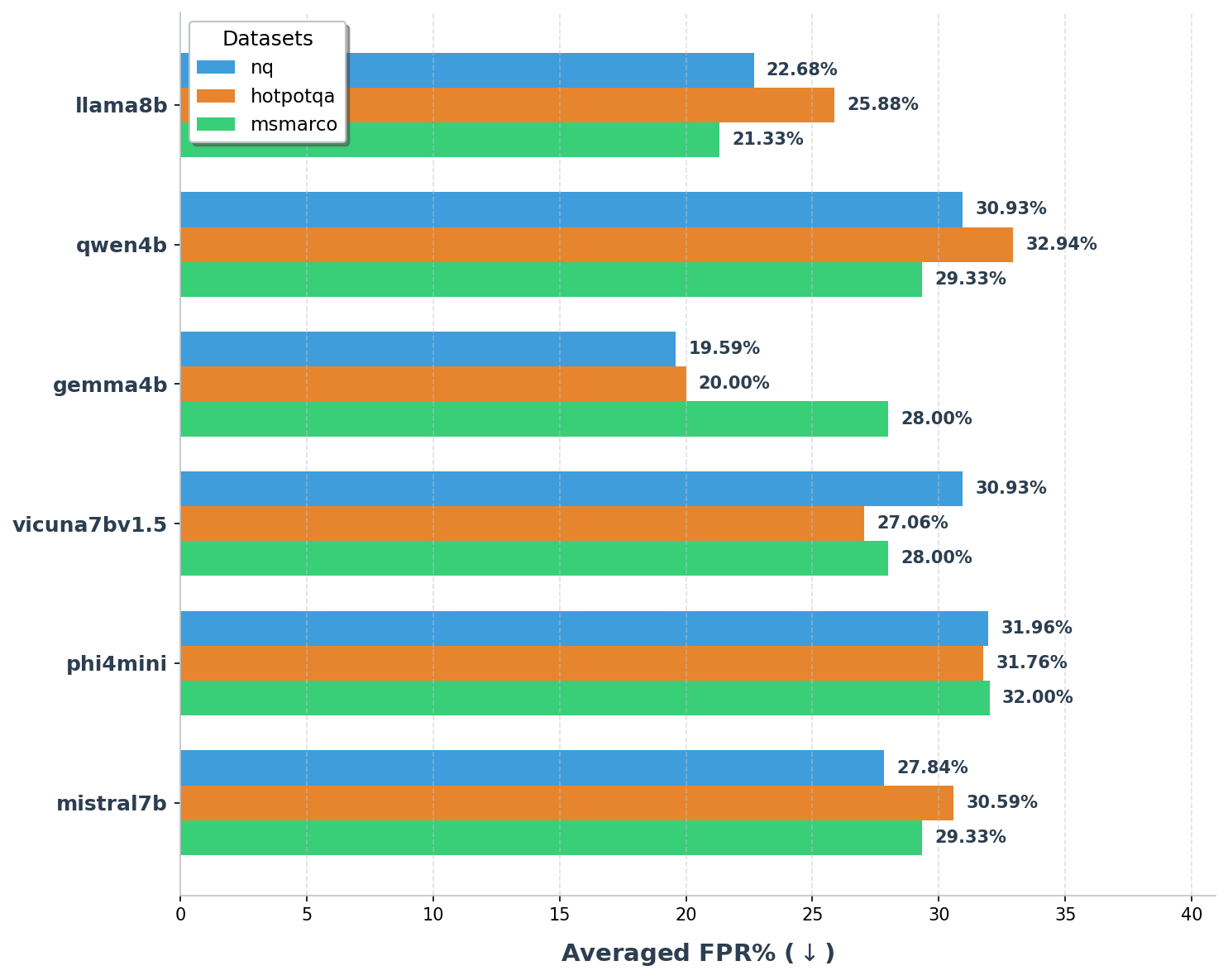}
    \caption{FPR results with $a_1=3$, $a_2=3$, $k_1=5$, $k_2=3$.}
    \label{appendix_result14}
\end{figure}

\begin{figure}[H]
    \centering    
    \includegraphics[width=0.48\textwidth]{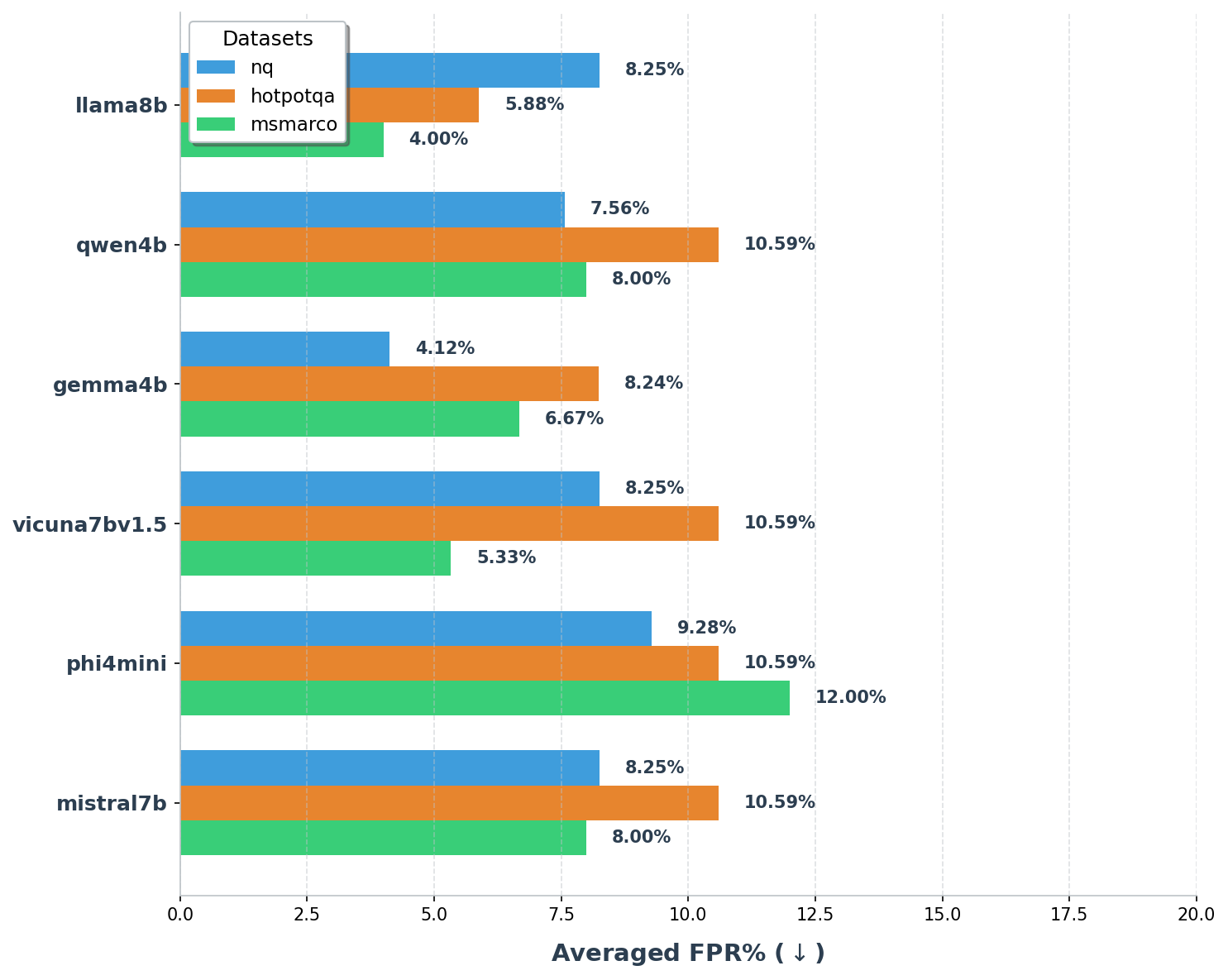}
    \caption{FPR results with $a_1=3$, $a_2=4$, $k_1=5$, $k_2=3$.}
    \label{appendix_result15}
\end{figure}

\begin{figure}[H]
    \centering    
    \includegraphics[width=0.48\textwidth]{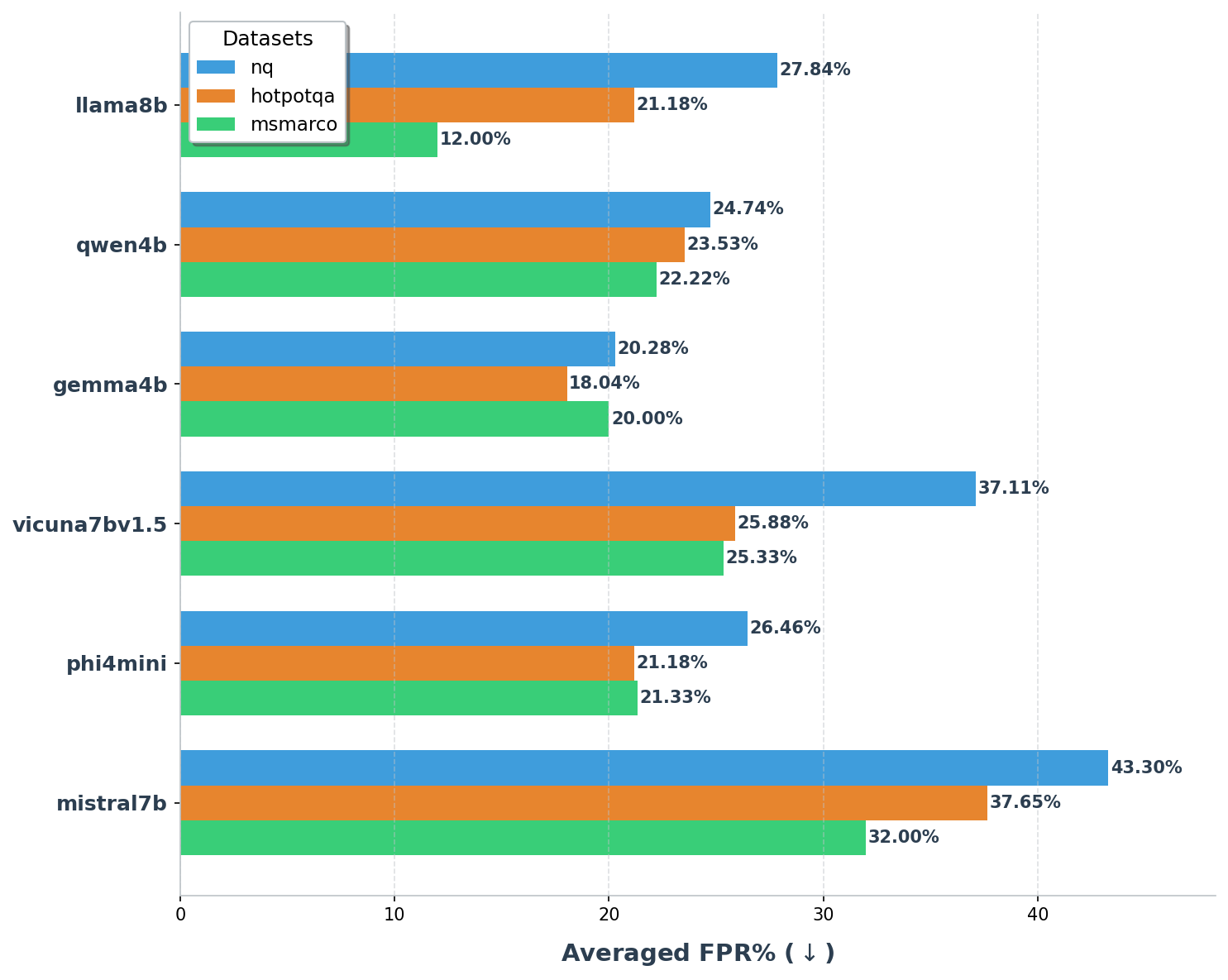}
    \caption{TPR results with $a_1=4$, $a_2=2$, $k_1=5$, $k_2=3$.}
    \label{appendix_result16}
\end{figure}

\subsection{ACC}

\begin{figure}[H]
    \centering    
    \includegraphics[width=0.48\textwidth]{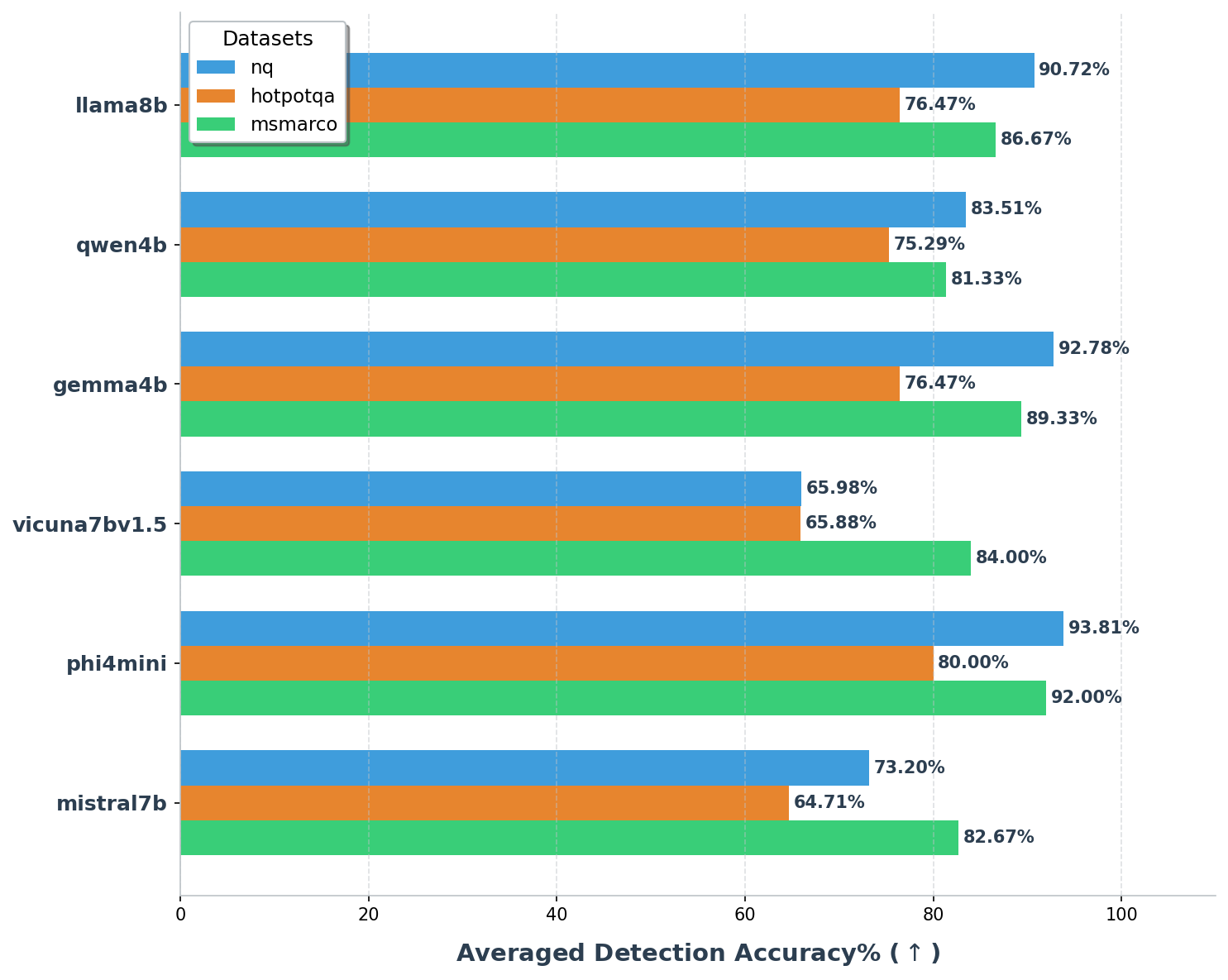}
    \caption{ACC results with $3$ retrieved documents, $a_1=2$, $a_2=2$, $k_1=5$, $k_2=3$.}
    \label{appendix_result17}
\end{figure}

\begin{figure}[H]
    \centering    
    \includegraphics[width=0.48\textwidth]{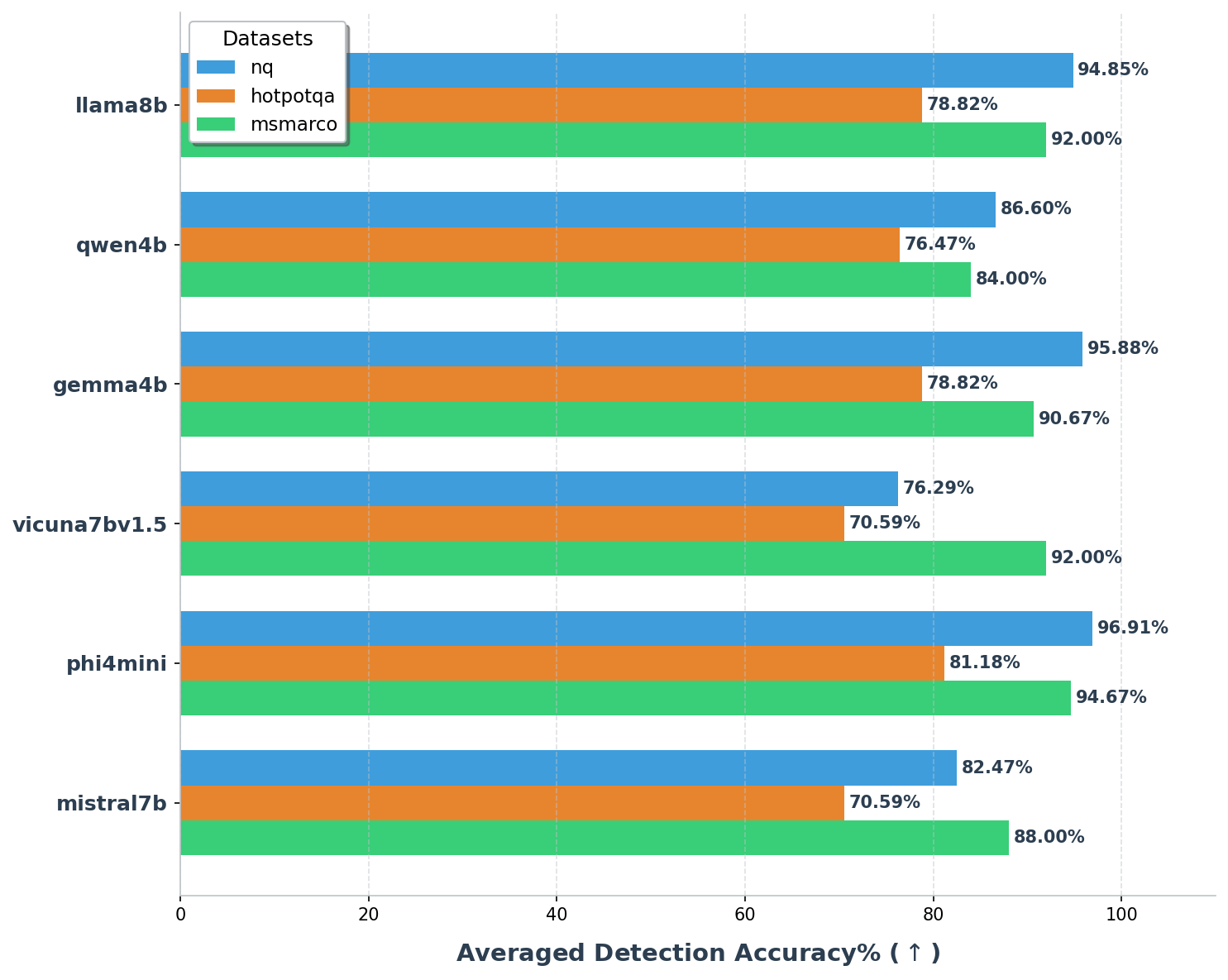}
    \caption{ACC results with $a_1=2$, $a_2=2$, $k_1=5$, $k_2=3$.}
    \label{appendix_result18}
\end{figure}

\begin{figure}[H]
    \centering    
    \includegraphics[width=0.48\textwidth]{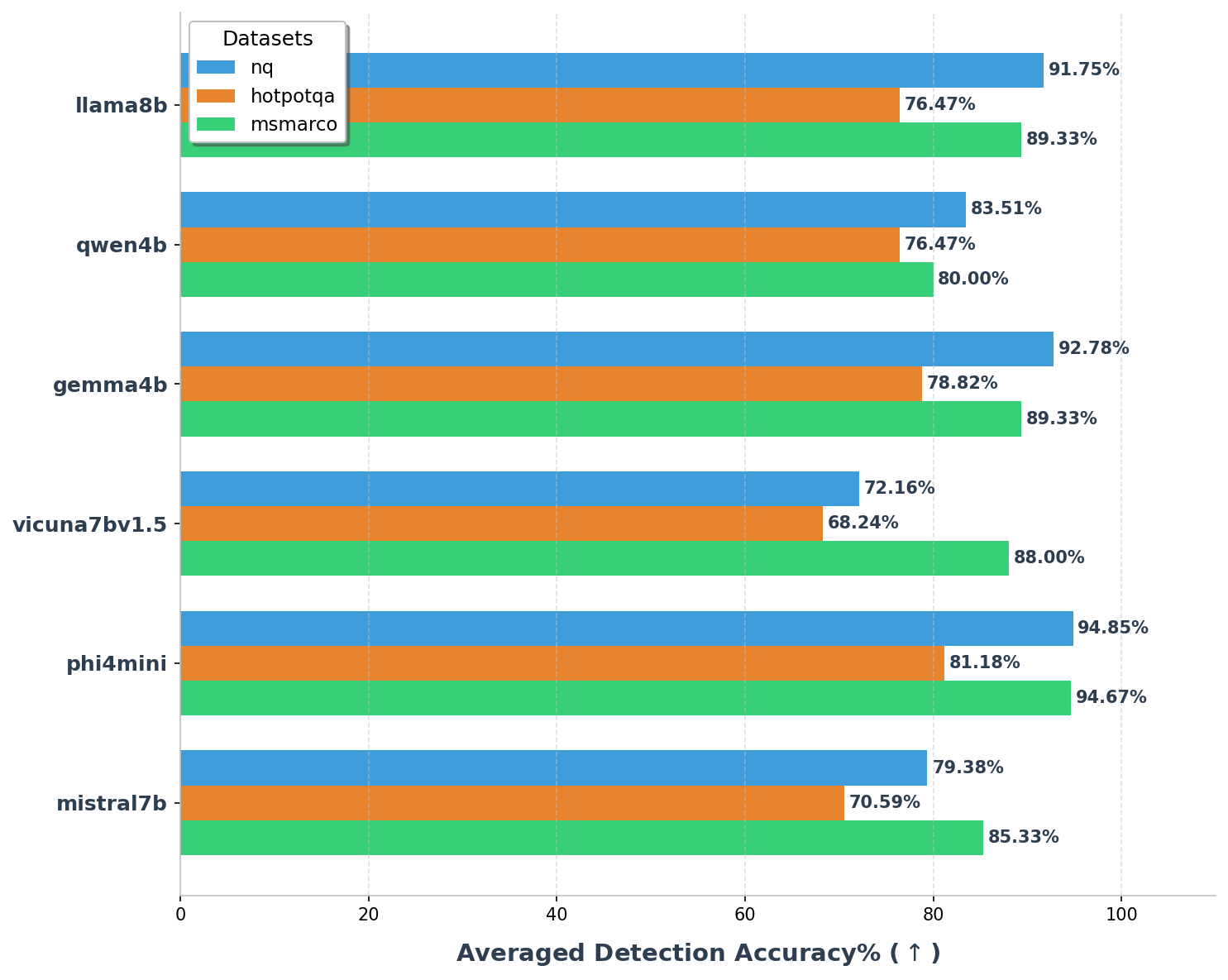}
    \caption{ACC results with $a_1=2$, $a_2=3$, $k_1=5$, $k_2=3$.}
    \label{appendix_result19}
\end{figure}

\begin{figure}[H]
    \centering    
    \includegraphics[width=0.48\textwidth]{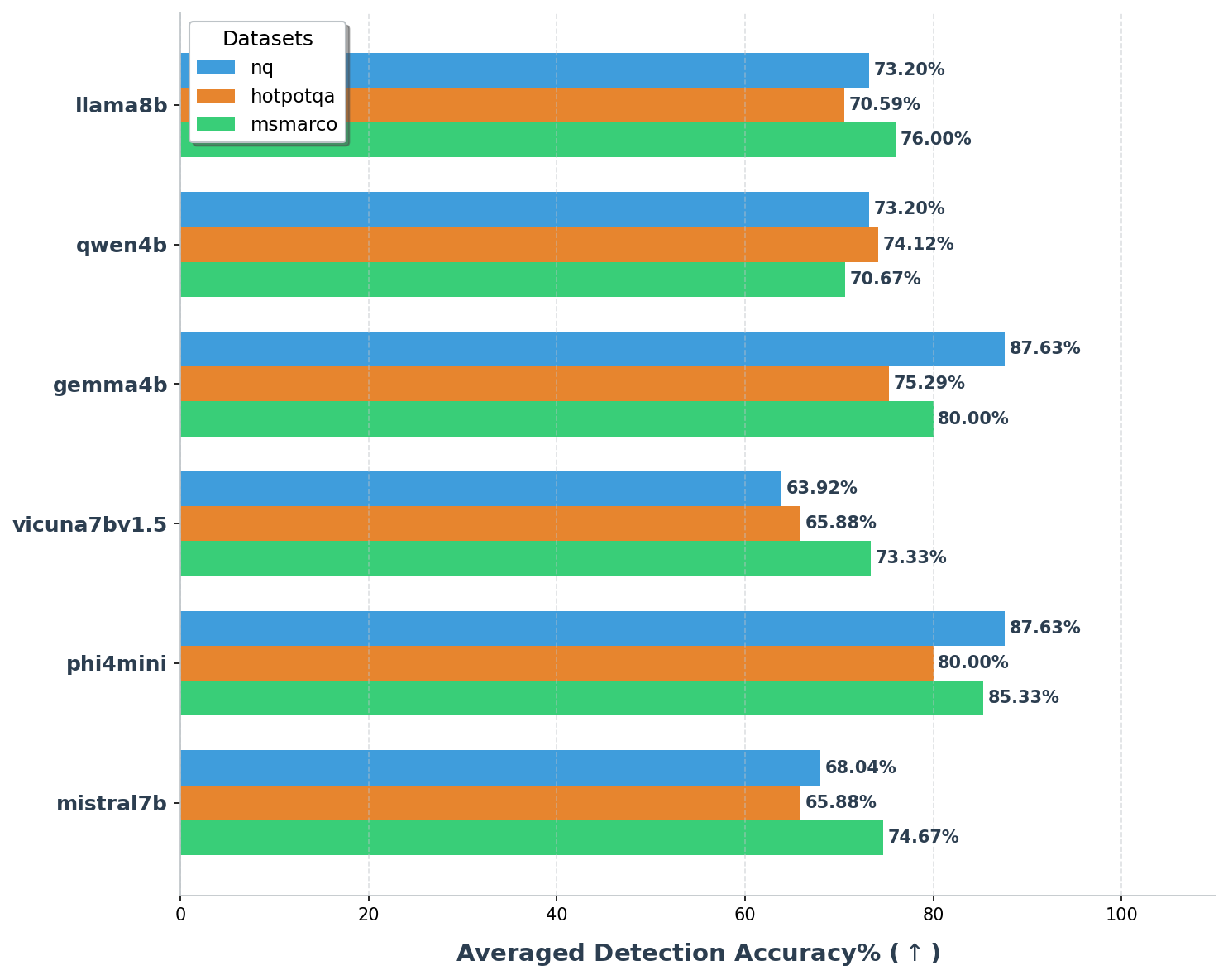}
    \caption{ACC results with $a_1=2$, $a_2=4$, $k_1=5$, $k_2=3$.}
    \label{appendix_result_20}
\end{figure}

\begin{figure}[H]
    \centering    
    \includegraphics[width=0.48\textwidth]{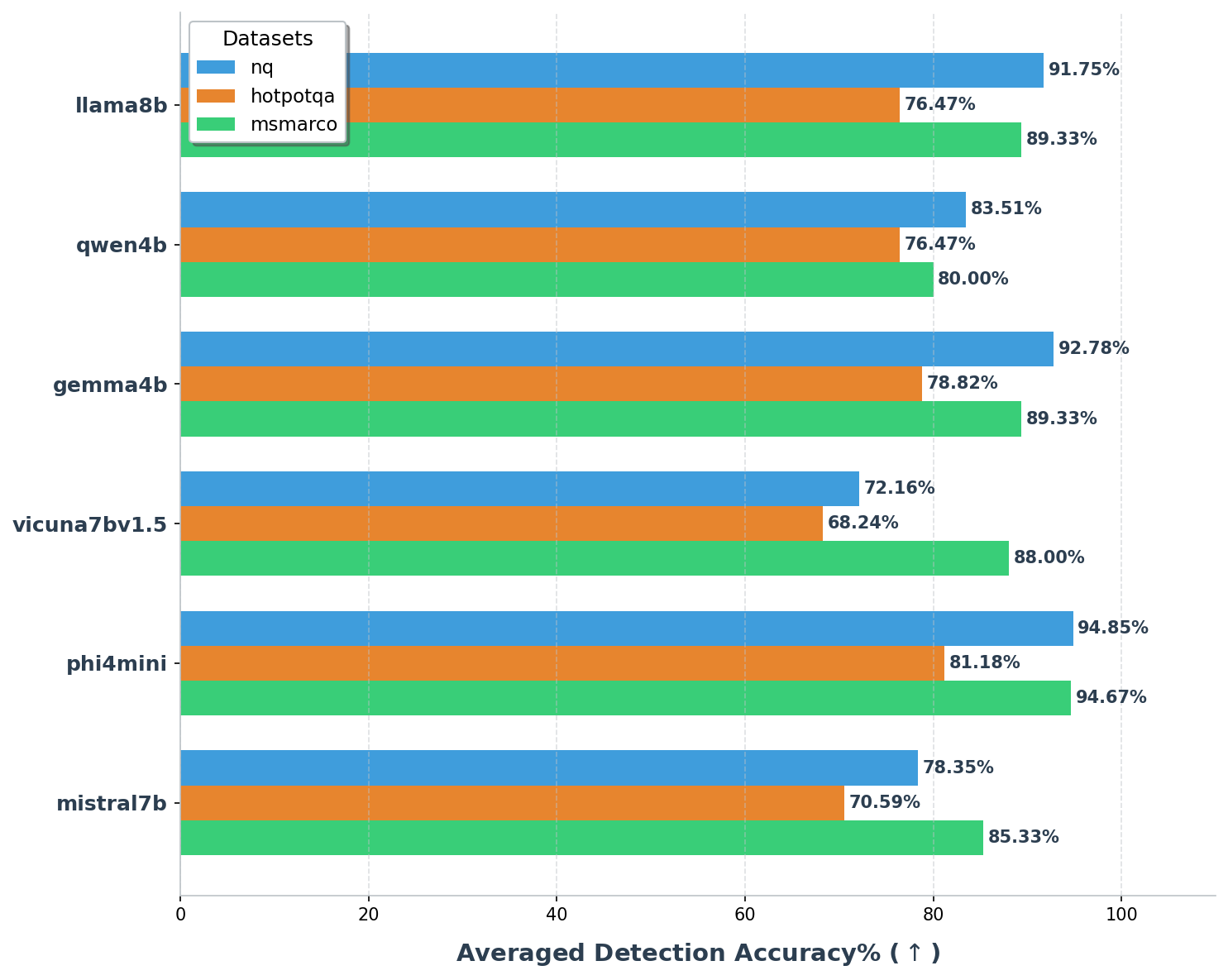}
    \caption{ACC results with $a_1=3$, $a_2=3$, $k_1=5$, $k_2=3$.}
    \label{appendix_result21}
\end{figure}

\begin{figure}[H]
    \centering    
    \includegraphics[width=0.48\textwidth]{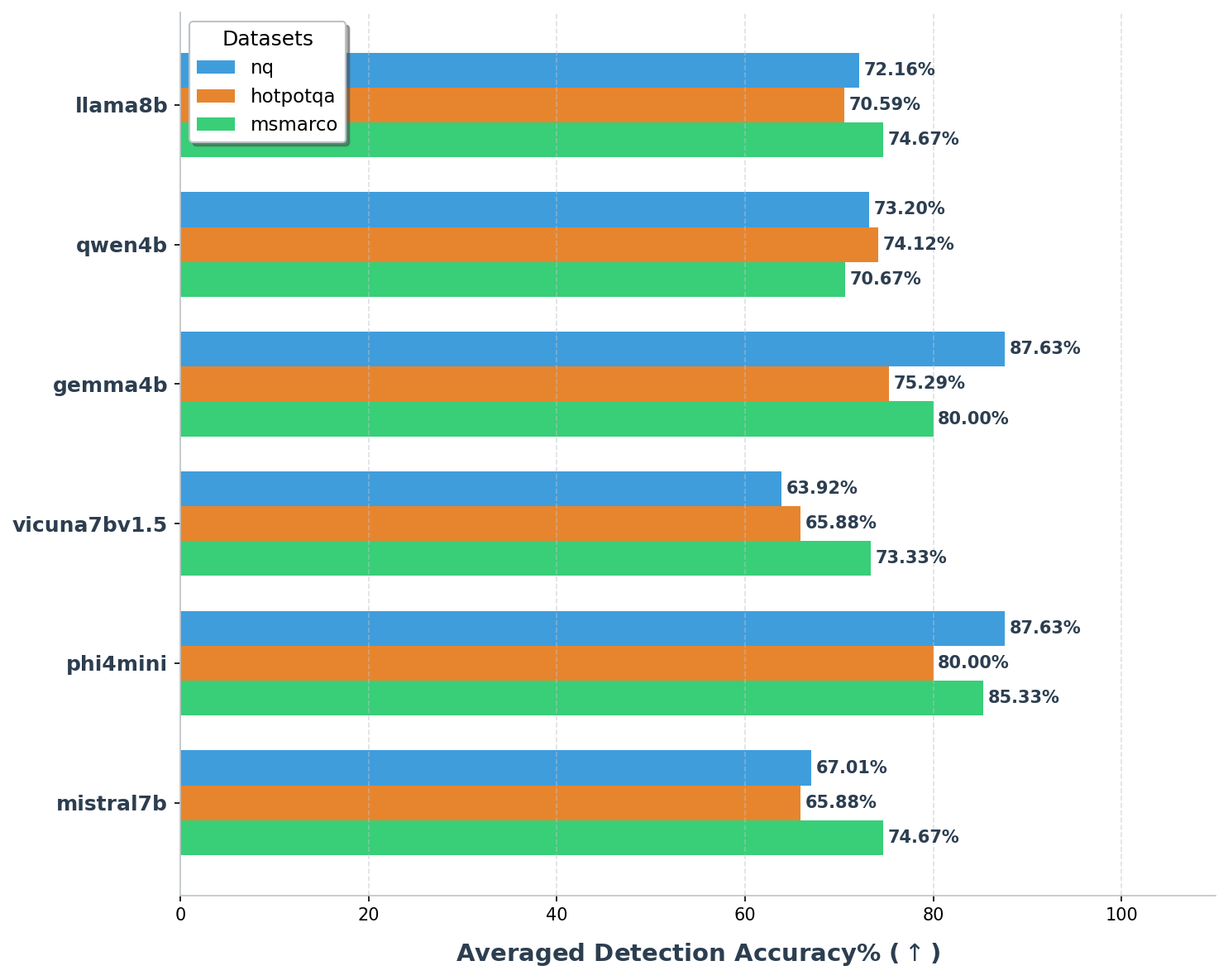}
    \caption{ACC results with $a_1=3$, $a_2=4$, $k_1=5$, $k_2=3$.}
    \label{appendix_result22}
\end{figure}

\begin{figure}[H]
    \centering    
    \includegraphics[width=0.48\textwidth]{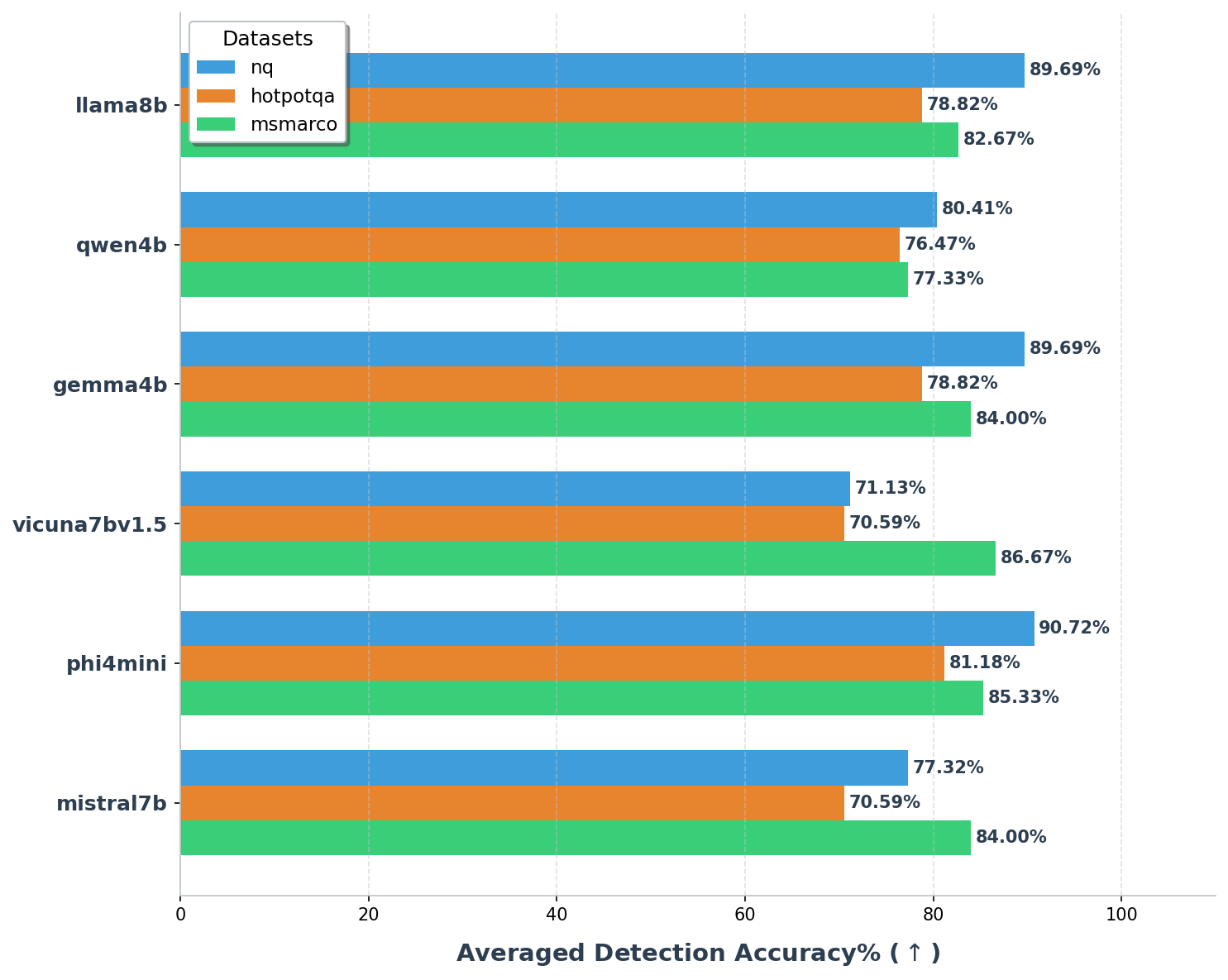}
    \caption{ACC results with $a_1=4$, $a_2=2$, $k_1=5$, $k_2=3$.}
    \label{appendix_result23}
\end{figure}

\begin{figure}[H]
    \centering    
    \includegraphics[width=0.48\textwidth]{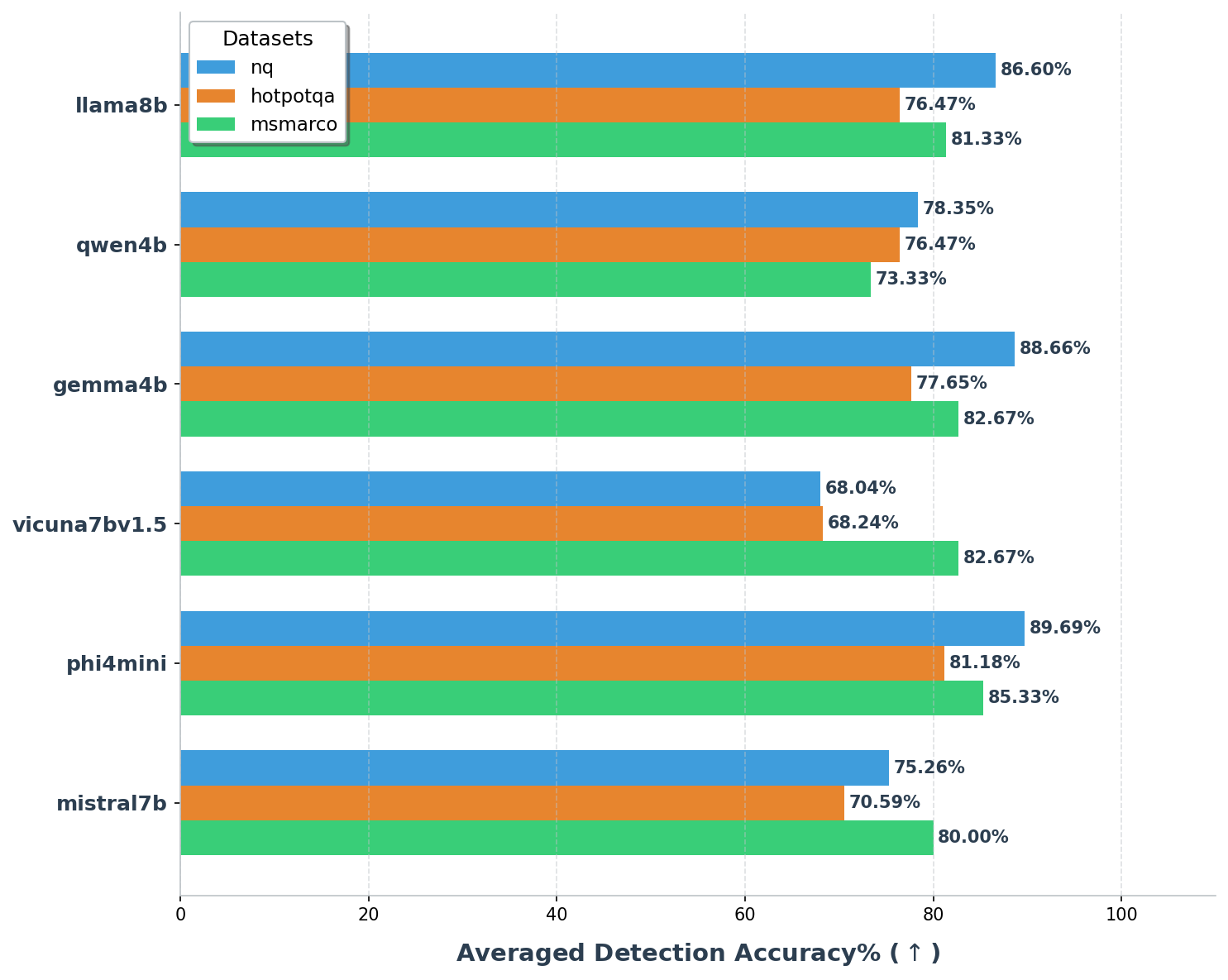}
    \caption{ACC results with $a_1=4$, $a_2=3$, $k_1=5$, $k_2=3$.}
    \label{appendix_result24}
\end{figure}

\begin{figure}[H]
    \centering    
    \includegraphics[width=0.48\textwidth]{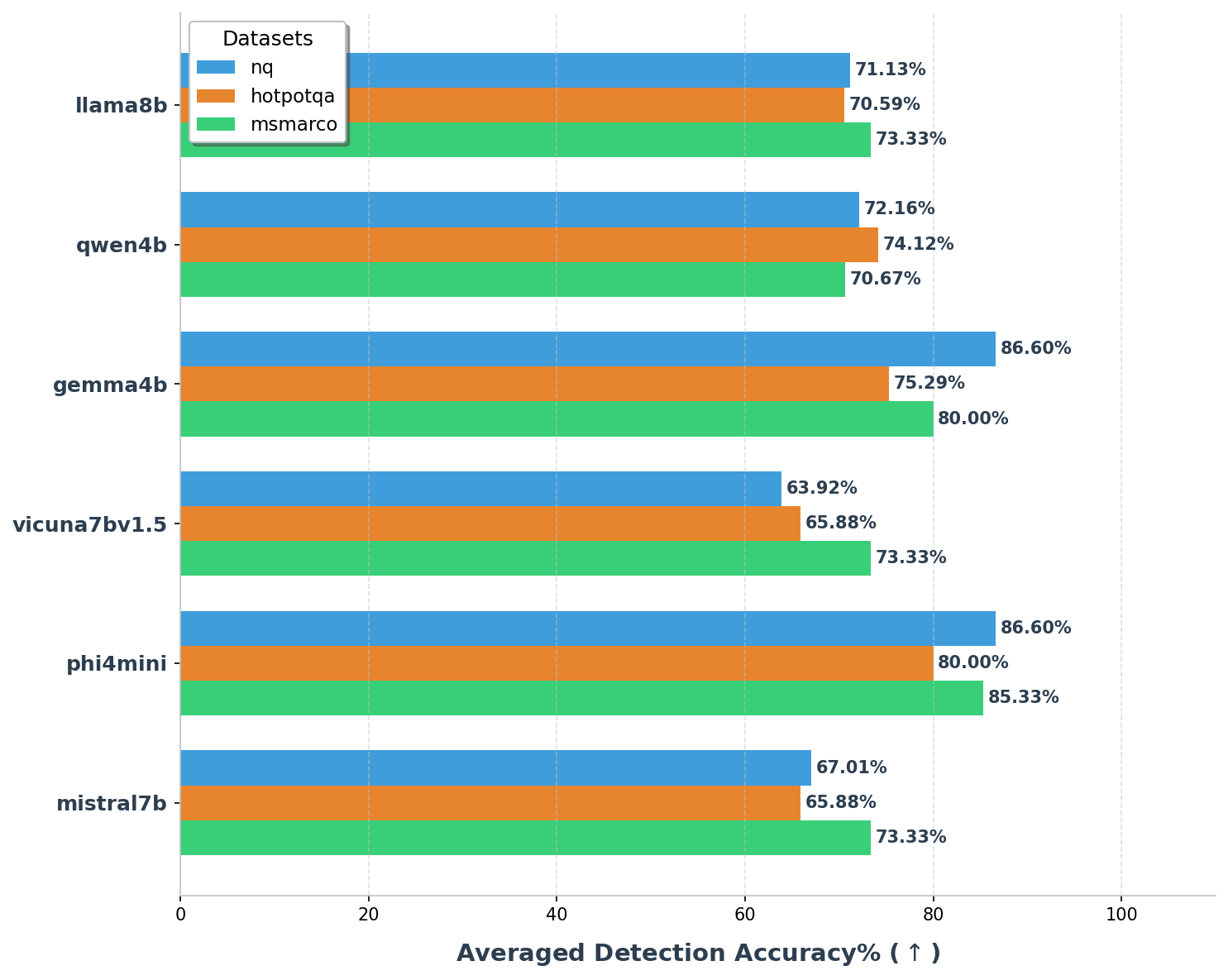}
    \caption{ACC results with $a_1=4$, $a_2=4$, $k_1=5$, $k_2=3$.}
    \label{appendix_result25}
\end{figure}

\end{document}